\pdfoutput=1
\documentclass[12pt]{article}

\usepackage{graphicx,psfrag,color}
\usepackage{amsmath,amssymb,amsfonts}
\usepackage{array}
\usepackage{cite}
\usepackage{rotating}

\newcounter{MBQ}

\newcommand{\grtsim}{\mbox{\raisebox{-3pt}{$\stackrel{>}{\sim}$}}}

\newcommand{\be}{\begin{equation}}
\newcommand{\ee}{\end{equation}}
\newcommand{\bea}{\begin{eqnarray}}
\newcommand{\eea}{\end{eqnarray}}
\newcommand{\bi}{\begin{itemize}}
	\newcommand{\ei}{\end{itemize}}
\newcommand{\ben}{\begin{enumerate}}
	\newcommand{\een}{\end{enumerate}}
\newcommand{\bt}{\begin{tabular}}
	\newcommand{\et}{\end{tabular}}

\begin{document}

%\allowdisplaybreaks
%\thispagestyle{empty}

\begin{titlepage}

\begin{flushright}
{\small
TUM-HEP-1065/16\\
FTUAM-16-38\\
IFT-UAM/CSIC-16-106\\
%arXiv:1611.0xxxx\\[0.0cm]
November 2, 2016
}
\end{flushright}

\vskip1cm
\begin{center}
{\large \bf 
The last refuge of mixed wino-Higgsino dark matter\\[0.1cm] 
}
\end{center}

\vspace{0.4cm}
\begin{center}
{\sc M.~Beneke$^{a}$, A.~Bharucha$^{b}$, A.~Hryczuk$^{c,d}$,} \\ 
{\sc  S.~Recksiegel$^{a}$, 
and  P. Ruiz-Femen\'\i a$^{a,e}$}\\[6mm]
{\it ${}^a$Physik Department T31,\\
James-Franck-Stra\ss e~1, 
Technische Universit\"at M\"unchen,\\
D--85748 Garching, Germany\\
\vspace{0.3cm}
${}^b$ Aix Marseille Univ, Universit\'{e} de Toulon, CNRS, CPT, Marseille, France\\
\vspace{0.3cm}
${}^c$Department of Physics, University of Oslo, Box 1048,\\ 
NO-0371 Oslo, Norway\\
\vspace{0.3cm}
${}^d$National Centre for Nuclear Research,\\
Ho\.za 69, 00-681, Warsaw, Poland\\
\vspace{0.3cm}
${}^e$Departamento de F\'{i}sica Te\'{o}rica and Instituto de F\'{i}sica Te\'{o}rica UAM-CSIC,\\
 Universidad Aut\'{o}noma de Madrid, E-28049 Madrid, Spain}
\\[0.3cm]
\end{center}

\vspace{0.7cm}
\begin{abstract}
\vskip0.2cm\noindent
We delineate the allowed parameter and mass range for a wino-like dark 
matter particle containing some Higgsino admixture in the MSSM by analysing 
the constraints from diffuse gamma-rays from the dwarf spheroidal galaxies, 
galactic cosmic rays, direct detection and cosmic microwave 
background an\-iso\-tro\-pies. A complete calculation of the Sommerfeld 
effect for the mixed-neutralino case is performed. We find that
the combination of direct and indirect searches poses significant restrictions 
on the thermally produced wino-Higgsino dark matter with correct relic 
density. For $\mu>0$ nearly the entire parameter space considered is 
excluded, while for $\mu<0$ a substantial region is still allowed, 
provided conservative assumptions on astrophysical uncertainties are adopted.
\end{abstract}
\end{titlepage}

%\setcounter{page}{1}
%\tableofcontents

%%%%%%%%%%%%%%%%%%%%%%%%%%%%%%%%%%%%%%%%%%%%%%%%%%%%%%%%%%%%%%%%%%%%%%%%%%%%%%
\section{Introduction}
\label{sec:introduction}

Many remaining regions in the parameter space of the Minimal Supersymmetric 
Standard Model (MSSM), which yield the observed thermal relic density for  
neutralino dark matter, rely on very specific mechanisms, such as 
Higgs-resonant annihilation in the so-called funnel region, or sfermion 
co-annihilation. In \cite{Beneke:2016ync} we identified new regions, where 
the dark matter particle is a mixed---as opposed to pure---wino, has mass 
in the TeV region, and yields the observed relic density. 
These new regions are driven to the correct relic abundance by the 
proximity of the resonance of the Sommerfeld effect due to electroweak gauge
boson exchange. In such situations, the annihilation cross section is 
strongly velocity dependent, and the present-day annihilation cross section 
is expected to be relatively large, potentially leading to observable signals 
in indirect searches for dark matter (DM). On the other hand, 
a substantial Higgsino fraction of a mixed dark matter particle leads  
to a large, potentially observable dark matter-nucleon scattering 
cross section. 

In this paper we address the question of which part of this region survives 
the combination of direct and indirect detection constraints. 
For the latter we consider diffuse gamma-rays from the dwarf 
spheroidal galaxies (dSphs), galactic cosmic rays (CRs) and cosmic microwave 
background (CMB) an\-iso\-tro\-pies. These have been found to be the most 
promising channels for detecting or excluding the pure-wino DM 
model \cite{Hryczuk:2014hpa}. Stronger limits can be obtained only from the 
non-observation of the gamma-line feature and to a lesser extent from  
diffuse gamma-rays both originating in the Galactic Centre (GC).
Indeed, it has been shown \cite{Cohen:2013ama,Fan:2013faa}
that the pure-wino model is ruled out by the absence of an excess in these 
search channels, unless the galactic dark matter profile develops a core, 
which remains a possibility. Since the viability of wino-like DM is a question 
of fundamental importance, we generally adopt the weaker constraint in 
case of uncertainty, and hence we take the point of view that wino-like DM 
is presently not excluded by gamma-line and galactic diffuse gamma-ray 
searches. Future results from the \v{C}erenkov Telescope Array (CTA) are 
expected to be sensitive enough to resolve this issue 
(see e.g. \cite{Roszkowski:2014iqa,Lefranc:2016fgn}), 
and will either observe an excess in gamma-rays or exclude the dominantly 
wino DM MSSM parameter region discussed in the present paper.

Imposing the observed relic density as a constraint, the pure-wino DM 
model has no free parameters and corresponds to the limit of the MSSM 
when all other superpartner particles and non-standard Higgs bosons 
are decoupled. Departing from the pure wino in the MSSM introduces 
many additional dimensions in the MSSM parameter space and changes the 
present-day annihilation cross section, branching ratios (BRs) for particular 
primary final states, and the final gamma and CR spectra
leading to a modification of the limits. The tools for the precise 
computation of neutralino dark matter (co-) annihilation in the generic MSSM 
when the Sommerfeld enhancement is operative have been developed 
in~\cite{Beneke:2012tg,Hellmann:2013jxa,Beneke:2014gja} and applied to 
relic density computations in \cite{Beneke:2014hja,Beneke:2016ync}.
The present analysis is based on an extension of the code to 
calculate the annihilation cross sections for all exclusive two-body 
final states separately, rather than the inclusive cross section.

Further motivation for the present study is provided by the spectrum of 
the cosmic antiproton-to-proton ratio reported by the AMS-02 
collaboration \cite{Aguilar:2016kjl}, which appears to be somewhat harder 
than expected from the commonly adopted cosmic-ray propagation models.
In \cite{Ibe:2015tma} it has been shown that pure-wino DM can improve the 
description of this data. Although our understanding of the background 
is insufficient to claim the existence of a dark matter signal in 
antiprotons, it is nevertheless interesting to check whether the 
surviving  mixed-wino DM regions are compatible with antiproton data. 

The outline of this paper is as follows. In Section~\ref{sec:fluxes} 
we summarize the theoretical input, beginning with a description of the 
dominantly wino MSSM parameter region satisfying the relic-density 
constraint, then providing some details on the computation of the 
DM annihilation rates to primary two-body final states. The following 
Section~\ref{sec:methodology} supplies information about the implementation of 
the constraints from diffuse gamma-rays from the dSphs, galactic CRs, direct 
detection and the CMB, and the data employed for the analysis. The results 
of the indirect detection analysis are presented in 
Section~\ref{sec:results} as constraints in the plane of the two most 
relevant parameters of the MSSM, the wino mass parameter $M_2$ and 
$|\mu|-M_2$, where $\mu$ is the 
Higgsino mass parameter. In Section~\ref{sec:maximalplot} the indirect 
detection constraints are combined with that from the non-observation 
of dark matter-nucleon scattering. 
For the case of $\mu<0$ we demonstrate the existence of a mixed wino-Higgsino 
region satisfying all constraints, while for $\mu>0$ we show that there is 
essentially no remaining parameter space left.
Section~\ref{sec:conclusions} concludes.
%%%%%%%%%%%%%%%%%%%%%%%%%%%%%%%%%%%%%%%%%%%%%%%%%%%%%%%%%%%%%%%%%%%%%%%%%%%%%%

\section{CR fluxes from wino-like dark matter}
\label{sec:fluxes}

\subsection{Dominantly-wino DM with thermal 
relic density in the MSSM}
\label{sec:mssm}

In \cite{Beneke:2016ync} the Sommerfeld corrections to the relic abundance 
computation for TeV-scale neutralino dark matter in the full MSSM have been 
studied. The ability to perform the computations for mixed dark matter 
at a general MSSM parameter space 
point \cite{Beneke:2012tg,Hellmann:2013jxa,Beneke:2014gja,Beneke:2014hja} 
revealed a large neutralino mass range with the correct thermal relic 
density, which opens mainly due to the proximity of the resonance of the 
Sommerfeld effect and its dependence on MSSM parameters. In this subsection 
we briefly review the dominantly-wino parameter region identified 
in \cite{Beneke:2016ync}, which will 
be studied in this paper. ``Dominantly-wino'' or ``wino-like'' here refers 
to a general MSSM with non-decoupled Higgs bosons, sfermions, bino and 
Higgsinos as long as the mixed neutralino dark matter state is mainly wino. 
We also require that its mass is significantly larger than the electroweak 
scale. The well-investigated pure-wino model refers to the limit in this 
parameter space, when all particles other than the triplet wino are decoupled. 

Despite the large number of parameters needed to specify a particular MSSM  
completely, in the dominantly-wino region, the annihilation rates depend  
strongly only on a subset of parameters. These are the wino, bino and 
Higgsino mass parameters $M_2$, $M_1$ and $\mu$, respectively, which control 
the neutralino composition and the chargino-neutralino mass difference, and 
the common sfermion mass parameter $M_{\rm sf}$. In this work we assume that 
the bino is much heavier that the wino, that is, the lightest neutralino 
is a mixed wino-Higgsino. Effectively a value of $|M_1|$ larger 
than $M_2$ by a few 100 GeV is enough to decouple the bino in the TeV 
region.\footnote{Allowing for significant bino admixture leads to other 
potentially interesting, though smaller regions, as described 
in \cite{Beneke:2016ync}.} The wino mass parameter determines the lightest 
neutralino (LSP) mass, and the difference $|\mu| - M_2$ 
the wino-Higgsino admixture. In the range $M_2=1-5$~TeV considered here,
the relation $m_{\rm LSP}\simeq M_2$ remains accurate to a few GeV, when 
some Higgsino fraction is added to the LSP state, and values of 
$|\mu|-M_2~\grtsim~500$~GeV imply practically decoupled Higgsinos. 

\begin{figure}[t]
\centering
\includegraphics[width=.59\textwidth]{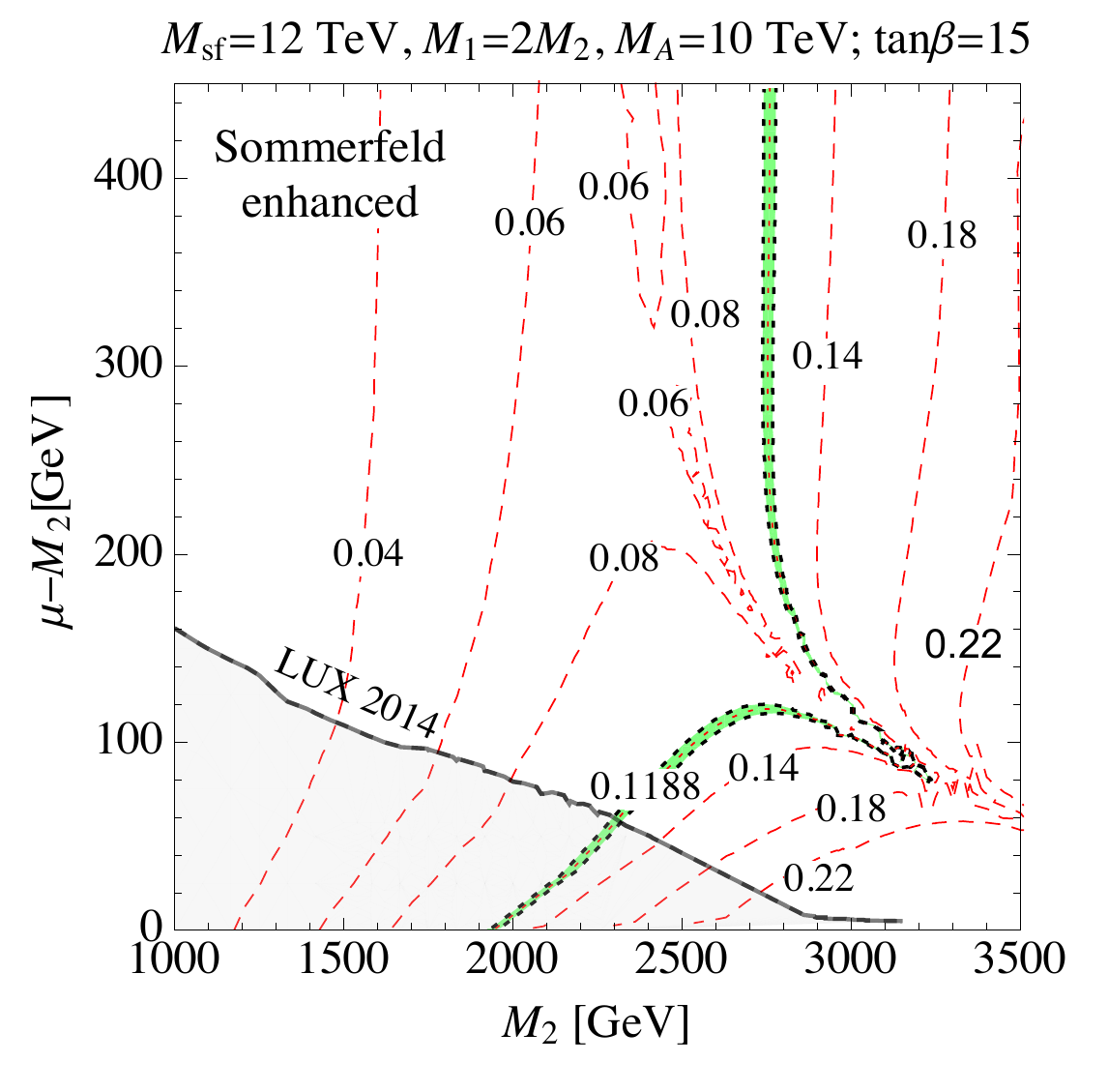}
\caption{Contours of constant relic density in the $M_2$ vs.\ $(\mu-M_2)$
plane for $\mu>0$, as computed in 
\cite{Beneke:2016ync}. The (green) band indicates the region within $2\sigma$ 
of the observed dark matter abundance. Parameters are as given in the 
header, and the trilinear couplings are set to $A_i= 8$~TeV for all sfermions 
except for that of the stop, which is fixed by the Higgs mass value. The 
black solid line corresponds to the old LUX limit~\cite{luxSI} on the 
spin-independent DM-nucleon cross section, which excludes the shaded area 
below this line. Relaxing the old LUX limit by a factor of two to account 
for theoretical uncertainties eliminates the direct detection constraint 
on the shown parameter space region. \label{fig:oldLUX}}
\end{figure}

Increasing the Higgsino component of the wino-like LSP lowers its coupling 
to charged gauge bosons, to which 
wino-like neutralinos annihilate predominantly, and therefore increases the 
relic density. Larger mixings also imply that the mass difference  between 
the lightest chargino and neutralino increases, which generically reduces 
the size of the Sommerfeld enhancement of the annihilation cross section. 
These features are apparent in the contours of constant relic density in the 
$|\mu|-M_2$ vs.\ $M_2$ plane for the wino-Higgsino case shown 
in \cite{Beneke:2016ync}, which are almost straight for large $|\mu|-M_2$, but 
bend to lower values of $m_{\rm LSP}$ as $|\mu|-M_2$ is reduced. 
A representative case is reproduced in  Fig.~\ref{fig:oldLUX}.
The contours also 
bend towards lower $M_2$ when sfermions become lighter, as they mediate 
the t- and u-channel annihilation into SM fermions, which interferes 
destructively with the s-channel annihilation, effectively lowering the 
co-annihilation cross section. By choosing small values of $M_{\rm sf}$ 
(but larger than $1.25~m_{\rm LSP}$ to prevent sfermion co-annihilation, 
not treated by the present version of the code), 
LSP masses as low as $1.7$~TeV are seen to give the correct thermal density, 
to be compared with the pure-wino result, $m_{\rm LSP}\simeq 2.8$~TeV.  

For $M_2 > 2.2$~TeV a resonance in the Sommerfeld-enhanced rates is present, 
which extends to larger $M_2$ values as the Higgsino fraction is increased. 
The enhancement of the cross section in the vicinity of the  resonance  makes 
the contours of constant relic density cluster around it and develop a peak 
that shifts $m_{\rm LSP}$ to larger values. In particular, the largest value 
of $M_2$, which gives the correct thermal relic density, is close to 3.3~TeV, 
approximately 20\% higher than for the pure-wino scenario. The influence of 
the less relevant MSSM Higgs mass parameter $M_A$ is also noticeable 
when the LSP contains 
some Higgsino admixture, which enhances the couplings to the Higgs (and Z) 
bosons in s-channel annihilation. This is more pronounced if $M_A$ is light enough 
such that final states containing heavy Higgs bosons are kinematically accessible. The corresponding increase 
in the annihilation cross section results in positive shifts of around 
100 to 250~GeV in the value of $M_2$ giving the correct relic density on 
decreasing $M_A$ from 10~TeV to 800~GeV. In summary, a 
large range of lightest neutralino masses, $1.7 - 3.5$~TeV, provides the 
correct relic density for the mixed wino-Higgsino state as a consequence of 
the Sommerfeld corrections.

The MSSM parameter points considered in this paper have passed standard 
collider, flavour and theoretical constraints as discussed 
in~\cite{Beneke:2016ync}. In the dominantly-wino parameter space, most of 
the collider and flavour constraints are either satisfied automatically or 
receive MSSM corrections that are suppressed or lie within the 
experimental and theoretical uncertainties. Ref.~\cite{Beneke:2016ync} further 
required compatibility with direct dark matter detection constraints by 
imposing that the DM-nucleon spin-independent cross section was less than 
twice the LUX limits reported at the time of publication~\cite{luxSI}.
This did not affect the results significantly, see Fig.~\ref{fig:oldLUX},  
as in most of the parameter space of interest the  
scattering cross section was predicted to be much above those limits.
Recently the LUX collaboration has presented a new limit, 
stronger than the previous one by approximately a factor of four
\cite{Akerib:2016vxi}, potentially imposing more severe constraints
on the dominantly-wino neutralino region of the MSSM parameter space.
The details of the 
implementation of the limits from indirect detection searches for the 
mixed wino, which were not included in our previous analysis, and  
from the new LUX results are given in Section~\ref{sec:methodology}.

%%%%%%%%%%%%%%%%%%%%%%%%%%%%%%%%%%%%%%%%%%%%%%%%%%%%%%%%%%%%%%%%%%%%%%%%%%%%%%
%%%%%%%%%%%%%%%%%%%%%%%%%%%%%%%%%%%%%%%%%%%%%%%%%%%%%%%%%%%%%%%%%%%%%%%%%%%%%%

%%%%%%%%%%%%%%%%%%%%%%%%%%%%%%%%%%%%%%%%%%%%%%%%%%%%%%%%%%%%%%%%%%%%%%%%%%%%%%
\subsection{Branching fractions and primary spectra}
\label{sec:model}

The annihilation of wino-like DM produces highly energetic particles, 
which subsequently decay, fragment and hadronize into stable SM 
particles, producing the CR fluxes. 

The primary particles can be any of the SM particles, and the heavy MSSM 
Higgs bosons, $H^0, A^0$ and $H^\pm$, when they are kinematically accessible. 
We consider neutralino dark matter annihilation into two primary particles. 
The number of such exclusive two-body channels is 31, and the corresponding
neutralino annihilation cross sections are computed including Sommerfeld 
loop corrections to the annihilation amplitude as described in 
\cite{Beneke:2016ync,Beneke:2012tg,Beneke:2014gja}. As input for this 
calculation we need to provide the tree-level exclusive annihilation rates 
of {\em all} neutral neutralino and chargino pairs, since through 
Sommerfeld corrections the initial LSP-LSP state can make transitions to 
other virtual states with heavier neutralinos or a pair of conjugated 
charginos, which subsequently annihilate into the primaries. 
The neutralino and chargino tree-level annihilation rates in the 
MSSM have been derived analytically in~\cite{Beneke:2012tg}, and 
including $v^2$-corrections in~\cite{Hellmann:2013jxa}, in the form of 
matrices, where the off-diagonal entries refer to the interference 
of the short-distance annihilation amplitudes of different 
neutralino/chargino two-particle states into the same final state. For the
present analysis the annihilation 
matrices have been generalized to vectors of matrices, such 
that the components of the vector refer to the 31 exclusive final 
states. The large number of different exclusive final states can be 
implemented without an increase in the CPU time for the computation 
relative to the inclusive case. Since  the information about the exclusive 
annihilation rates only enters through the
(short-distance) annihilation matrices, the two-particle wave-functions 
that account for the (long-distance) Sommerfeld corrections  only
need to be computed once. On the contrary, since the $v^2$-corrections 
to the annihilation of DM in the present Universe are very small, 
they can be neglected, which results in a significant reduction in the 
time needed to compute the annihilation matrices.\footnote{Since we 
also computed the relic density for every parameter point, which requires 
including the $v^2$-corrections, we did not 
make use of this simplification in the present analysis.} It further 
suffices to compute the present-day annihilation cross section for 
a single dark matter velocity, and we choose $v=10^{-3}\,c$. The reason 
for this choice is that the Sommerfeld effect saturates for very small 
velocities, and the velocity dependence is negligible for velocities smaller 
than $10^{-3}\,c$.

The energy spectrum $d N_f/dx$ of a stable particle $f$ at 
production per DM annihilation can be written as
\begin{align}
\label{eq:spectra}
\frac{d N_f}{dx} = \sum_I {\rm Br}_I \, \frac{d N_{I\to f}}{dx}\,, 
\end{align}
where $x=E_f/m_{\rm LSP}$, and $d N_{I\to f}/dx$ represents the contribution
from each two-body primary final state $I$ with branching fraction 
${\rm Br}_I$ to the spectrum of $f$ after the decay, fragmentation and 
hadronization processes have taken place. We compute ${\rm Br}_I$ from 
our MSSM Sommerfeld code as described above and use the tables 
for $d N_{I\to f}/dx$ provided with the PPPC4DMID code~\cite{Cirelli:2010xx}, 
which include the leading logarithmic electroweak corrections through the 
electroweak fragmentation functions~\cite{Ciafaloni:2010ti}.

Two comments regarding the use of the spectra provided by the PPPC4DMID 
code are in order. The code only considers
primary pairs $I$ of a particle together with its antiparticle, both assumed 
to have the same energy spectrum. 
For wino-like DM there exist primary final states with 
different species, {\it i.e.} $I=ij$ with $j\ne \bar{i}$, such as $Z\gamma$ 
and $Zh^0$. In this case, we compute the final number of particles $f$ 
produced from that channel as one half of the sum of those produced by 
channels $I=i\bar{i}$ and $I=j\bar{j}$. This is justified, since the 
fragmentation of particles $i$ and $j$ is independent. 
A second caveat concerns the heavy MSSM Higgs bosons that can be
produced for sufficiently heavy neutralinos. These are not considered to be 
primary channels in the PPPC4DMID code, which only deals with SM particles.
A proper treatment of these primaries would first account for the 
decay modes of the heavy Higgs bosons, and then 
consider the fragmentation and hadronization of the SM multi-particle 
final state in an event generator. Instead of a full treatment, we  
replace the charged Higgs $H^\pm$ by a longitudinal-polarized 
$W^\pm$-boson, and the neutral heavy Higgses $H^0,\, A^0$ 
by the light Higgs $h^0$ when computing the spectra in $x$. This 
approximation is not very well justified. However, the branching 
ratios of the dominantly-wino neutralino to final states 
with heavy Higgses are strongly suppressed, and we could equally have 
set them to zero without a noticeable effect on our results. 

\begin{figure}[t]
  \centering
  \includegraphics[width=.49\textwidth]{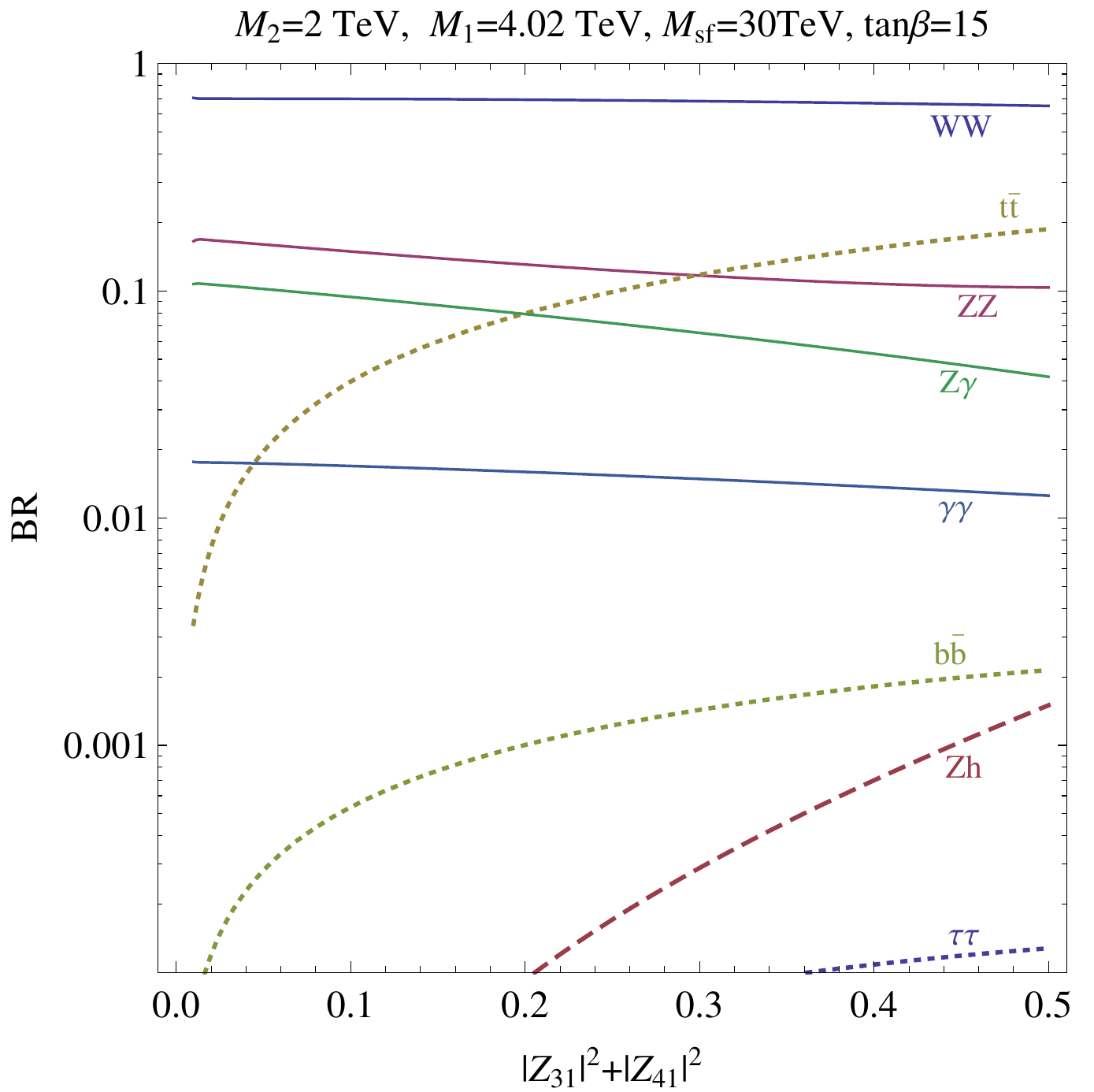}
  \includegraphics[width=.49\textwidth]{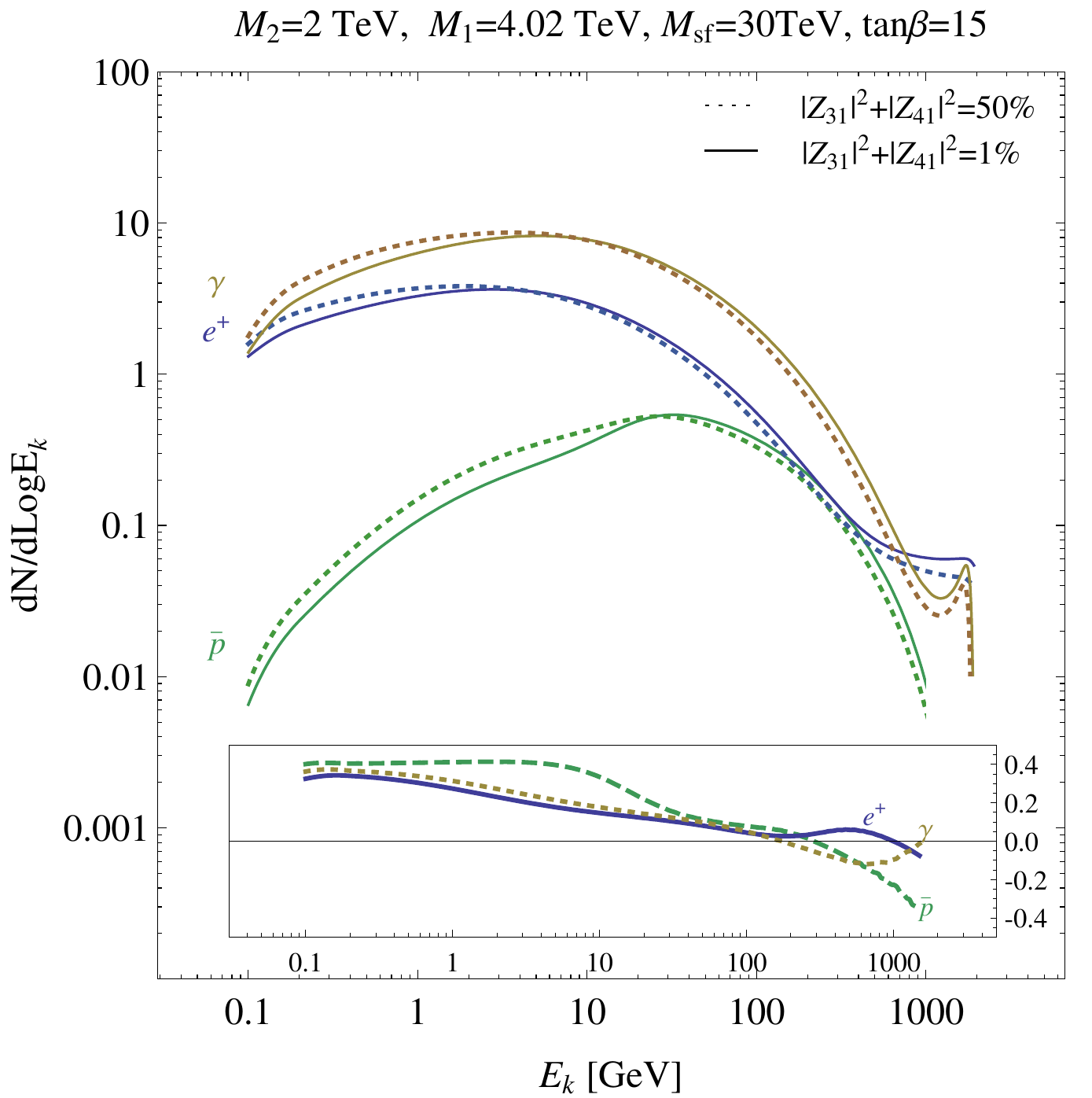}
\caption{Left: Branching fractions of present-day wino-like 
neutralino annihilation vs.\ the Higgsino fraction for decoupled $M_A$ 
and sfermions. $|Z_{31}|^2+|Z_{41}|^2$ refers to the Higgsino fraction of 
the lightest neutralino in the convention of \cite{Beneke:2012tg}.
Right: Comparison of $\bar p$, $e^+$ and gamma-ray spectra 
per annihilation at production of a 50\% mixed wino-Higgsino (dashed)
to the pure-wino (solid) model. The gamma-line component is not  
shown. In the inset at the bottom of the plot the relative differences between 
the two spectra are shown.\label{fig:fluxes}}
\end{figure}

The branching fractions of primary final states obtained from our code are 
shown in the left panel of Fig.~\ref{fig:fluxes} as a function of the 
Higgsino fraction for a wino-like LSP with 2~TeV mass. The 
pure wino annihilates mostly to $W^+W^-$ and to a lesser extent to  
other pairs of gauge bosons, including the loop-induced photon final 
state, which is generated by the Sommerfeld correction. The annihilation to 
fermions is helicity or $p$-wave suppressed. The suppression is lifted only 
for the $t\bar{t}$ final state as the Higgsino admixture increases, in which 
case this final state becomes the second most important. Except for this 
channel, the dominant branching fractions are largely independent of the 
Higgsino fraction. The annihilation to $W^+W^-$ is always 
dominant and above 75\%.  

The final spectra of photons, positrons and antiprotons per annihilation at 
production for small (solid lines) and large (dashed lines) Higgsino mixing
are plotted in the right panel of Fig.~\ref{fig:fluxes}. The spectra in these
two extreme cases are very similar, because $W^+ W^-$ is the dominant 
primary final state largely independent of the wino-Higgsino composition, 
and also the number of final stable particles produced by the sub-dominant 
primary channels do not differ significantly from each other. The inset 
in the right-hand plot shows that the relative change between the mixed and 
pure wino case varies from about $+40\%$ to about $-40\%$ over the 
considered energy range. Concerning the variation with respect to the DM 
mass, the most important change is in the total annihilation cross section, 
not in the spectra $d N_f/dx$. The branching ratios ${\rm Br}_I$ to 
primaries depend on the LSP mass in the TeV regime only through the Sommerfeld 
corrections, which can change the relative size of the different channels. 
However, since for wino-like neutralinos annihilation into $W^+W^-$ dominates the sum over $I$ in  (\ref{eq:spectra}), the dependence of the final 
spectra on $m_{\rm LSP}$ is very mild. 

%%%%%%%%%%%%%%%%%%%%%%%%%%%%%%%%%%%%%%%%%%%%%%%%%%%%%%%%%%%%%%%%%%%%%%%%%%%%%%
%%%%%%%%%%%%%%%%%%%%%%%%%%%%%%%%%%%%%%%%%%%%%%%%%%%%%%%%%%%%%%%%%%%%%%%%%%%%%%

%%%%%%%%%%%%%%%%%%%%%%%%%%%%%%%%%%%%%%%%%%%%%%%%%%%%%%%%%%%%%%%%%%%%%%%%%%%%%%

\section{Indirect and direct searches}
\label{sec:methodology}

In this section we discuss our strategy for determining the constraints on 
mixed-wino dark matter from various indirect searches. While 
the analysis follows 
that for the pure wino \cite{Hryczuk:2014hpa}, here we focus on the most 
relevant search channels: the diffuse gamma-ray emission from dSphs, 
antiprotons and positron CRs, and the CMB. Moreover, since we consider 
wino-like DM with a possibly significant Higgsino admixture, we implement 
the direct detection constraints as well.

%%%%%%%%%%%%%%%%%%%%%%%%%%%%%%%%%%%%%%%%%%%%%%%%%%%%%%%%%%%%%%%%%%%%%%%%%%%%%%

\subsection{Charged cosmic rays}
\label{sec:CRs}

\subsubsection{Propagation}
\label{sec:propagation}

The propagation of charged CRs in the Galaxy is best described within the 
diffusion model with possible inclusion of convection. In this 
framework the general propagation equation takes the form 
\cite{Ginzburg:1990sk}
\begin{eqnarray}
\label{eq:propag}
&&\frac{\partial N^i}{\partial t}-
\vec{\nabla}\cdot\left(D_{xx} \vec{\nabla}-\vec{v}_c\right) N^i+
\frac{\partial}{\partial p}\left(\dot p-\frac{p}{3}\,
\vec{\nabla}\cdot \vec{v_c}\right) N^i -
\frac{\partial}{\partial p}p^2 D_{pp} \frac{\partial}{\partial p}
\frac{N^i}{p^2} 
\\
&&= Q^i(p,r,z)+\sum_{j>i} c\beta n_{\rm gas}(r,z)\sigma_{ij}N^j - 
c\beta n_{\rm gas}(r,z)\sigma_{{\rm in}} N^i -
\sum_{j<i} \frac{N^{i}}{\tau^{i \rightarrow j}} + 
\sum_{j>i} \frac{N^{j}}{\tau^{j \rightarrow i}}\;, 
\nonumber
\end{eqnarray}
where $N^i(p,r,z)$ is the number density of the $i$-th particle species 
with momentum $p$ and corresponding velocity $v=c\beta$, written in 
cylindrical coordinates $(r,z)$, 
$\sigma_{\rm in}$ the inelastic scattering cross section, $\sigma_{ij}$ the 
production cross section of species $i$ by the fragmentation of species $j$, 
and $\tau^{i \rightarrow j}$, $\tau^{j \rightarrow i}$ are the lifetimes 
related to decays of $i$ and production from heavier species $j$, 
respectively.

We solve \eqref{eq:propag} with the help of the DRAGON 
code \cite{Evoli:2008dv}, assuming cylindrical symmetry and no convection,  
$\vec{v}_c=0$. With the galacto-centric radius $r$, 
the height from the Galactic disk $z$ and rigidity $R=pc/Ze$, we adopt the
following form of the spatial diffusion coefficient:
\begin{equation}
D_{xx}(R,r,z)=D_0 \beta^\eta\left(\frac{R}{R_0}\right)^{\!\delta} 
e^{|z|/z_d}e^{(r-r_{\odot})/r_d}.
\label{eq:diffcoeff}
\end{equation}
The momentum-space diffusion coefficient, also referred to as reaccelaration, 
is related to it via $D_{pp}D_{xx}=p^2 v_A^2/9$, where the Alfv\'{e}n 
velocity $v_A$ represents the characteristic velocity of a 
magnetohydrodynamic wave. The free parameters are the normalization $D_0$, 
the spectral indices 
$\eta$ and $\delta$, the parameters setting the radial scale $r_d$ and 
thickness $z_d$ of the diffusion zone, and finally $v_A$. We fix the 
normalization at $R_0=3$~GV. The diffusion coefficient is assumed to 
grow with $r$, as the large scale galactic magnetic field gets weaker far 
away from the galactic center.

The source term is assumed to have the form 
\begin{equation}
Q^i(R,r,z)=f^i(r,z)\left(\frac{R}{R^i}\right)^{\!-\gamma^i},
\label{eq:source}
\end{equation}
where $f^i(r,z)$ parametrizes the spatial distribution of supernova remnants 
normalized at $R^i$, and $\gamma^i$ is the injection spectral index 
for species $i$. For protons and Helium we modify the source term to 
accommodate for two breaks in the power-law, as strongly indicated by  
observations. Leptons lose energy very efficiently, thus those which are very energetic
need to be very local, while we do not observe nor expect many local 
sources of TeV scale leptons. This motivates multiplying~\eqref{eq:source} 
by an additional exponential cut-off in energy, $e^{-E/E_c}$, with $E_c$ 
set to 50 TeV for electron and positron injection spectra.

We employ the gas distribution $n_{\rm gas}$ derived 
in \cite{Tavakoli:2012jx,Pohl:2007dz} 
and adopt the standard force-field approximation \cite{Gleeson:1968zza} to 
describe the effect of solar modulation. The modulation potential is assumed 
to be a free parameter of the fit and is allowed to be different for 
different CR species.

\subsubsection{Background models}
\label{sec:Bckgr}

%%%%%%%%%%%%%%%%%%%%%%%%%%%%%%%%%%%%%%%%%%%%%%%%%%%%%%%%%%%%%%%%%%%%%%%%%%
{\tabcolsep=0.16cm
\begin{table}
\centering
\footnotesize{
\begin{tabular}{|c c |c  c c c|c c |c c|}
\hline \multicolumn{ 2}{|c|}{Benchmark} & \multicolumn{ 4}{c|}{Diffusion} & 
\multicolumn{ 4}{c|}{Injection} 
\\ \cline{ 1- 10}
 Model & $z_d$ & $\delta$  & $D_0/10^{28}$  & $v_A$ & $\eta$ & 
$\gamma^p_1/\gamma^p_2/\gamma^p_3$ & $R^p_{0,1}$ & 
$\gamma^{He}_1/\gamma^{He}_2/\gamma^{He}_3$ & $R^{He}_{0,1}$ \\ 
& [kpc]  &  &  [${\rm cm}^2{\rm s}^{-1}$] &  [${\rm km\; s}^{-1}$] & & & GV 
& & GV  \\ 
\hline
 Thin & 1 & 0.47  & 0.43  & 13.0 & $-0.37$ & $1.85/2.39/2.22$ & 6.8 & 
$2.18/2.35/2.16$ & 12.7   \\ 
\hline
 Med & 4 & 0.5  & 1.79 & 14.0 & $-0.3$ & $1.90/2.36/2.21$ & 7.7 & 
$2.18/2.35/2.16$ & 12.7    \\ 
\hline
 Thick & 10 & 0.5  & 3.3 & 14.5 & $-0.27$ & $2.00/2.38/2.22$ & 7.5  & 
$2.18/2.36/2.12$ & 12.7    \\ 
\hline
\end{tabular}
}
\caption{Benchmark propagation models. The radial length is always $r_d=20$ 
kpc and convection is neglected, $\vec{v}_c=0$. The second break in the proton 
injection spectra is at 300 GV. For primary electrons we use a broken 
power-law with spectral indices $1.6/2.65$ and a break at 7 GV, while for 
heavier nuclei we assumed one power-law with index 2.25. $R^i_{0,1}$ 
refer to the positions of the first and second break, respectively, 
and $\gamma^i_{1,2,3}$ to the power-law in the three regions separated 
by the two breaks. The propagation 
parameters were obtained by fitting to B/C, proton and He data and 
cross-checked with antiproton data, while the primary electrons were 
obtained from the measured electron flux.}
\label{tab:propagation}
\end{table}
}
%%%%%%%%%%%%%%%%%%%%%%%%%%%%%%%%%%%%%%%%%%%%%%%%%%%%%%%%%%%%%%%%%%%%%%%%%%

\begin{figure}[t]
  \centering
  \includegraphics[width=.49\textwidth]{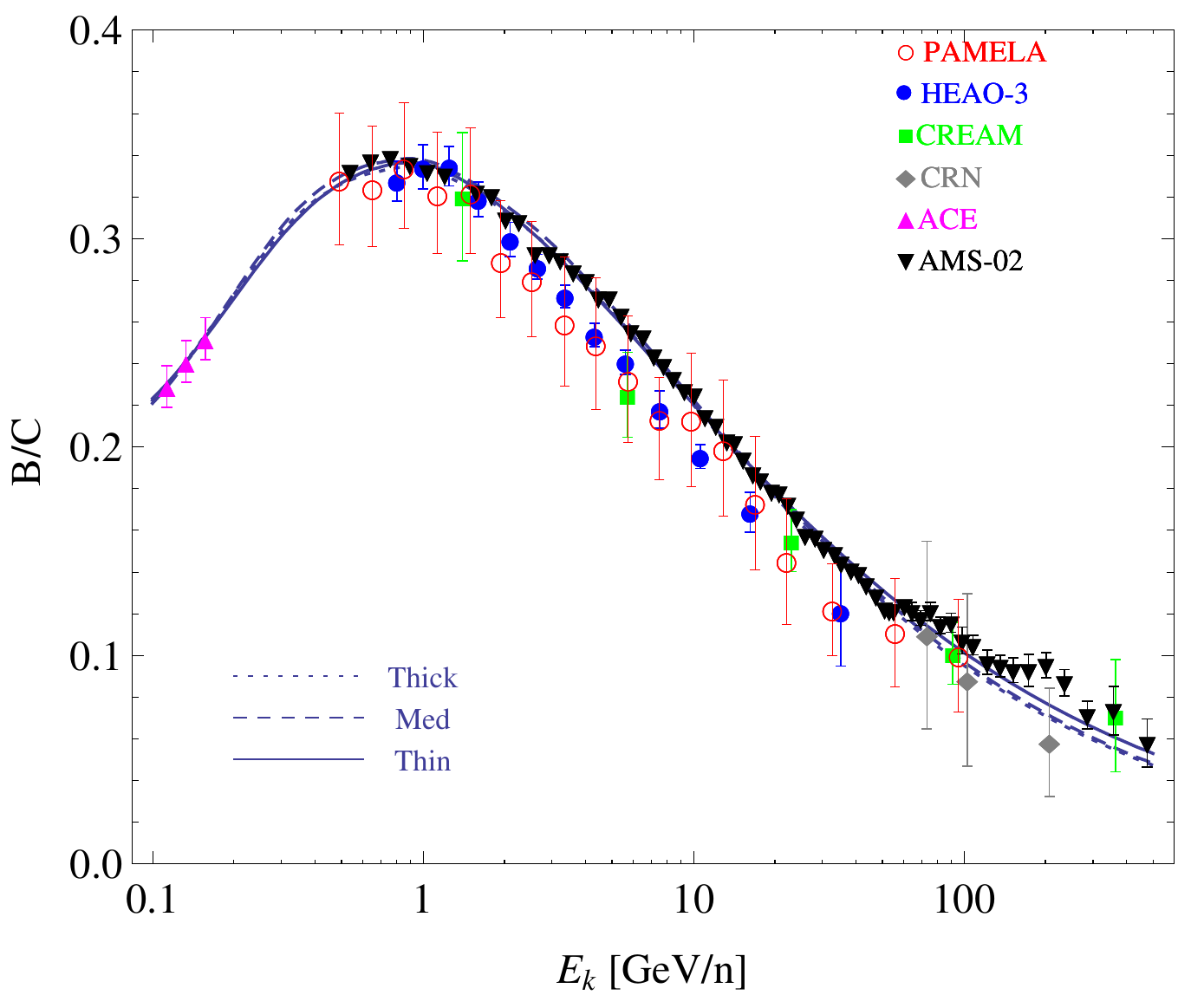}
  \includegraphics[width=.49\textwidth,height=6.6cm]{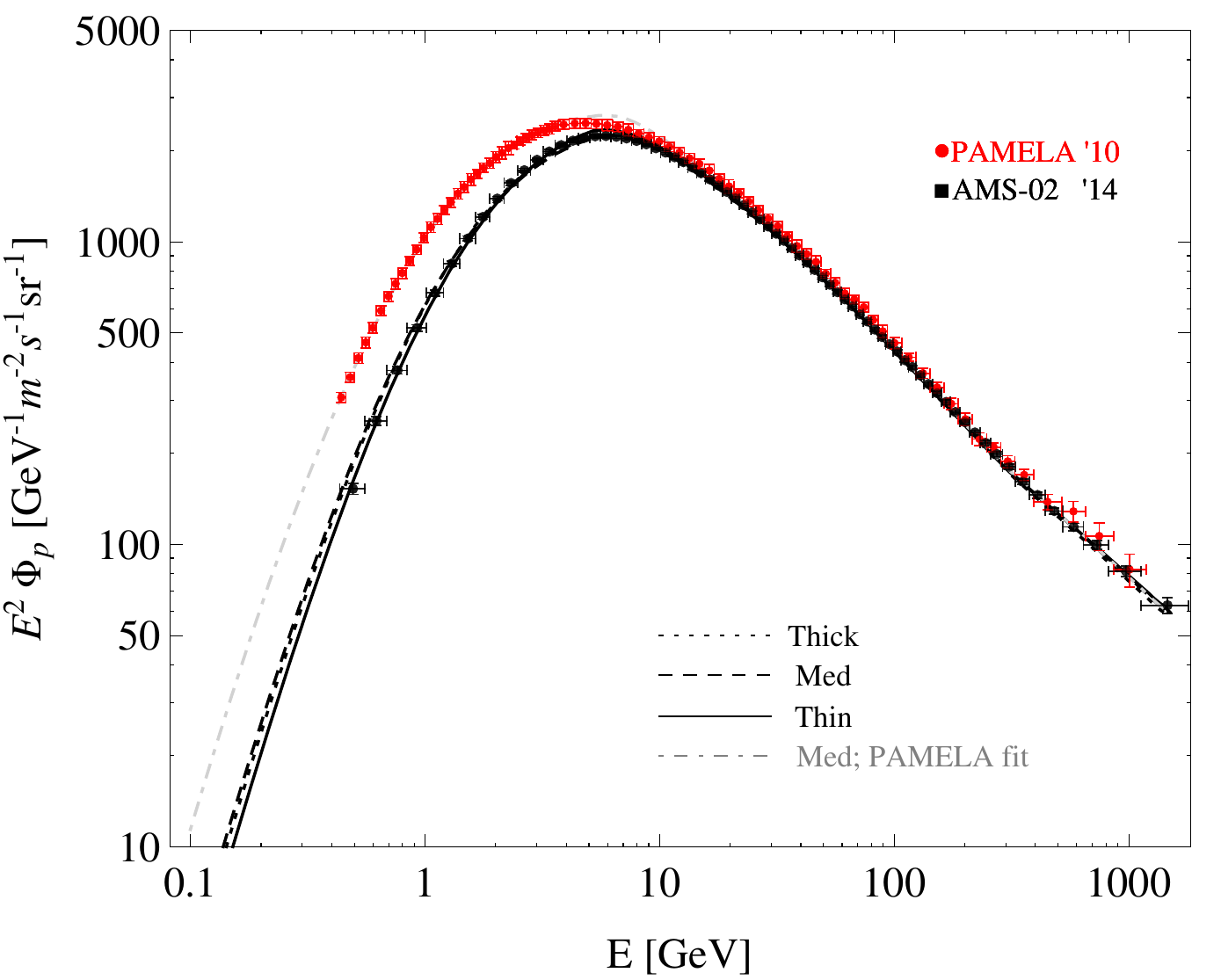}
\caption{Comparison of the benchmark propagation models: B/C (left) and 
protons (right). The fit was performed exclusively to the AMS-02 \cite{AMSdays2,AMSdays3,Aguilar:2015ooa}
measurements, while the other data sets are shown only for comparison: PAMELA \cite{Adriani:2011cu,Adriani:2014xoa},
HEAO-3 \cite{Engelmann:1990zz}, CREAM \cite{Ahn:2008my}, CRN \cite{1990ApJ.349.625S}, ACE \cite{0004-637X-698-2-1666}.
\label{fig:CRs}}
\end{figure}

In \cite{Hryczuk:2014hpa} 11 benchmark propagation models with 
varying diffusion zone thickness, from $z_d=1$ kpc to $z_d=20$ kpc, 
were identified by fitting to the B/C, proton, Helium, electron and 
$e^++e^-$ data. Since then the AMS-02 experiment provided 
CR spectra with unprecedented precision, which necessitates 
modifications of the above benchmark models. Following the same procedure 
as in \cite{Hryczuk:2014hpa} we choose three representative models, which 
give a reasonable fit to the AMS-02 data, 
denoted Thin, Med and Thick, corresponding 
to the previous $z_d=1$~kpc, $z_d=4$~kpc and $z_d=10$~kpc 
models.\footnote{We loosely follow here the widely adopted MIN, MED, MAX 
philosophy \cite{Donato:2003xg}, choosing models with as large variation 
in the DM-originated antiproton flux as possible. However, the MIN, MED, MAX 
models were optimized for pre-AMS data and are based on a semi-analytic 
diffusion model. Since we rely on the full numerical solution of the diffusion 
equation, we follow the benchmark models of \cite{Hryczuk:2014hpa}. This 
comes at the expense of no guarantee that the chosen models 
really provide the minimal and maximal number of antiprotons. However, as 
in this work we are not interested in setting precise \textit{limits} 
from antiproton data, we consider this approach as adequate.} 
The relevant parameters are given in Table \ref{tab:propagation}. In 
Fig.~\ref{fig:CRs} we show the fit to the B/C and the AMS-02 proton data 
\cite{AMSdays2,AMSdays3,Aguilar:2015ooa} 
and superimpose the older data from 
PAMELA \cite{Adriani:2011cu,Adriani:2014xoa}. In all these cases, as well as 
for the lepton data \cite{Accardo:2014lma,Aguilar:2014fea}, the measurements 
used in the fits were from AMS-02 results only. 

%%%%%%%%%%%%%%%%%%%%%%%%%%%%%%%%%%%%%%%%%%%%%%%%%%%%%%%%%%%%%%%%%%%%%%%%%%
%%%%%%%%%%%%%%%%%%%%%%%%%%%%%%%%%%%%%%%%%%%%%%%%%%%%%%%%%%%%%%%%%%%%%%%%%%

In the fit we additionally assumed that the normalization of the 
secondary CR antiprotons can freely vary by 10\% with respect to the result 
given by the DRAGON code. This is motivated by the uncertainty in the 
antiproton production cross sections. The impact of this and other 
uncertainties has been studied 
in detail in e.g. \cite{Kappl:2015bqa,Evoli:2015vaa,Giesen:2015ufa}.

As we will show below, the DM contribution to the lepton spectra is of much 
less importance for constraining the parameter space of our interest, 
therefore, we do not discuss the lepton backgrounds explicitly. All the 
details of the implementation of the lepton limits closely 
follow \cite{Hryczuk:2014hpa}, updated to 
the published AMS-02 data \cite{Accardo:2014lma,Aguilar:2014fea}.

%%%%%%%%%%%%%%%%%%%%%%%%%%%%%%%%%%%%%%%%%%%%%%%%%%%%%%%%%%%%%%%%%%%%%%%%%%%%%%

\subsection{Diffuse gamma-rays from dSphs}
\label{sec:dSphs}

Recently the \textit{Fermi}-LAT and MAGIC collaborations released  
limits from the combination of their   
stacked analyses of 15 dwarf spheroidal galaxies \cite{Ahnen:2016qkx}. 
Here we use the results of this analysis to constrain the parameter space of 
the mixed wino-Higgsino neutralino. To this end we compute all exclusive 
annihilation cross sections for present-day DM annihilation in the halo and 
take a weighted average of the limits provided by the experimental 
collaborations. As discussed in Section~\ref{sec:model}, the TeV scale 
wino-like neutralino annihilates predominantly into $W^+W^-$, $ZZ$ and 
$t\bar t$, with much smaller rates into leptons and the lighter 
quarks. In the results from \cite{Ahnen:2016qkx} only the $W^+W^-$, 
$b\bar b$, $\mu^+\mu^-$ and $\tau^+\tau^-$ final states are given. However, 
as the predicted 
spectrum and number of photons from a single annihilation is not significantly 
different for the hadronic or leptonic final states, we adopt the 
approximation that the limits from annihilation into $ZZ$ are the same as 
from $W^+W^-$, while those from $t\bar t$ and $c\bar c$ are the same as 
$b\bar b$. The differences in the number of photons produced between these 
annihilation channels in the relevant energy range are maximally of order 
$\mathcal{O}(20\%)$ for $W^+W^-$ vs.\ $ZZ$ and $t\bar t$ vs.\ $b\bar b$. 
Comparing $b\bar b$ to light quarks these can rise up to factor 2, however 
due to helicity suppression these channels have negligible branching 
fractions. Hence, the adopted approximation is expected to be  very good
and, the corresponding uncertainty is significantly smaller than that 
related to the astrophysical properties of the dSphs (parametrised by 
the $J$-factors).

%%%%%%%%%%%%%%%%%%%%%%%%%%%%%%%%%%%%%%%%%%%%%%%%%%%%%%%%%%%%%%%%%%%%%%%%%%%%%%

\subsection{CMB constraints}
\label{sec:CMB}
The annihilation of dark matter at times around recombination can affect the 
recombination history of the Universe by injecting energy into the 
pre-recombination photon-baryon plasma and into the post-recombination 
gas and background radiation, which has consequences for the power and 
polarization spectra of the CMB~\cite{Padmanabhan:2005es,Galli:2009zc,Slatyer:2009yq}.
In particular, it can result in the attenuation of the temperature and 
polarization power spectra, more so on smaller scales, and in a shift of the 
TE and EE peaks. These effects can be traced back to the increased ionization
fraction and baryon temperature, resulting in a broadening of the surface of 
last scattering, which suppresses perturbations on scales less than the width 
of this surface. 
Therefore the CMB temperature and polarization angular power spectra can be 
used to infer upper bounds on the annihilation cross section of dark matter 
into a certain final state for a given mass. When Majorana dark matter 
particles annihilate, the rate at which energy $E$ is released per unit 
volume $V$ can be written as
\begin{equation}
\frac{dE}{dt dV}(z)=\rho_{\rm crit}^2 \Omega^2(1+z)^6 p_{\rm ann}(z)
\end{equation} 
where $\rho_{\rm crit}$ is the critical density of the Universe today, and 
experiment provides constraints on $ p_{\rm ann}(z)$, which describes the 
effects of the DM. These effects are found to be well enough accounted for 
when the $z$ dependence of $p_{\rm ann}(z)$ is neglected, such that a limit is 
obtained for the constant $p_{\rm ann}$. The latest 95\% C.L. upper limit on 
$p_{\rm ann}$ was obtained by Planck~\cite{Ade:2015xua}, and we adopt their 
most significant limit $3.4\cdot 10^{-28}\,{\rm cm}^3 {\rm s}^{-1} 
{\rm GeV}^{-1}$ from the combination of TT, TE, EE~+~lowP~+~lensing data. 
The constant $p_{\rm ann}$ can further be expressed via
\begin{equation}
p_{\rm ann}=\frac{1}{M_{\chi}} f_{\rm eff} \langle \sigma v \rangle,
\end{equation}
where $f_{\rm eff}$, parametrizing the fraction of the rest mass energy that 
is injected into the plasma or gas, must then be calculated in order to 
extract bounds on the DM annihilation cross section in the recombination era.
In our analysis, for $f_{\rm eff}$ we use the quantities $f^I_{\rm eff, new}$ 
from~\cite{Madhavacheril:2013cna} for a given primary  annihilation channel 
$I$. We then extract the upper limit on the annihilation cross section at the 
time of recombination by performing a weighted average over the contributing 
annihilation channels, as done for the indirect detection limits discussed in 
Section~\ref{sec:dSphs}. As the Sommerfeld effect saturates before this time, 
$\langle \sigma v \rangle$ at recombination is the same as the present-day 
cross section. In the future the cross section bound can be improved by 
almost an order of magnitude, until $p_{\rm ann}$ is ultimately limited by 
cosmic variance.

\subsection{Direct detection}
\label{sec:DD}

Direct detection experiments probe the interaction of the dark matter 
particle with nucleons. For the parameter space of interest here, the bounds 
on spin-independent interactions, sensitive to the t-channel exchange of the 
Higgs bosons and to s-channel sfermion exchange are more constraining than 
those on spin-dependent interactions. The coupling of the lightest neutralino 
to a Higgs boson requires both a Higgsino and gaugino component, and is 
therefore dependent on the mixing. Note that the relative size of the Higgs 
Yukawa couplings means that the contribution due to the Higgs coupling  
to strange quarks dominates the result. 

In the pure-wino limit, when the sfermions are decoupled and the coupling to 
the Higgs bosons vanishes, the direct detection constraints are very weak as 
the elastic scattering takes place only at the loop level 
\cite{Hisano:2011cs}. Allowing for a Higgsino admixture and/or 
non-decoupled sfermions introduces tree-level scattering processes mediated 
by Higgs or sfermion exchange. Direct detection experiments have recently 
reached the sensitivity needed to measure such low scattering cross sections 
and with the new data released by the LUX \cite{Akerib:2016vxi} and  
PandaX \cite{Tan:2016zwf} collaborations, a portion of 
the discussed parameter space is now being probed.

In the analysis below we adopt the LUX limits \cite{Akerib:2016vxi}, being 
the strongest in the neutralino mass range we consider. In order to be 
conservative, in addition to the limit presented by the collaboration we 
consider a weaker limit by multiplying by a factor of two. This factor two 
takes into account the two dominant uncertainties affecting the 
spin-independent cross section, i.e.~the local relic density of dark matter 
and the strange quark content of the nucleon. The former, 
$\rho=0.3\pm 0.1~\mathrm{GeV}/\mathrm{cm}^3$, results in an uncertainty of 
$ 50\%$~\cite{Bovy:2012tw} and the latter result contributes an uncertainty 
on the cross section of about $20\%$~\cite{Durr:2015dna}, which on 
combination result in weakening the bounds by a factor of two (denoted as 
$\times 2$ on the plots). For the computation of the spin-independent 
scattering cross section for every model point we use micrOMEGAs 
\cite{Belanger:2010gh,Belanger:2013oya}. Note that 
the Sommerfeld effect does not influence this computation and the tree-level 
result is expected to be accurate enough.

Since only mixed Higgsino-gaugino neutralinos couple to Higgs bosons, the 
limits are sensitive to the parameters affecting the mixing. To be precise, 
for the case that the bino is decoupled ($|M_1|\gg M_2,|\mu|$) and 
$|\mu|-M_2 \gg m_Z$, the couplings of the Higgs bosons $h,H$ to the lightest 
neutralino are proportional to
\begin{equation}
c_{h}= m_Z c_W \frac{ M_2 + \mu\sin 2\beta }{\mu^2-M_2^2},
\qquad c_H =-m_Z c_W \frac{\mu \cos 2\beta }{\mu^2-M_2^2},
\end{equation}
where $c_W\equiv \cos\theta_W$, and it is further assumed that $M_A$ is 
heavy such that $c_{h,H}$ can be computed in the decoupling limit 
$\cos(\alpha-\beta)\to 0$. When $\tan\beta$ increases, the light Higgs 
coupling $c_h$ decreases for $\mu>0$ and increases for $\mu<0$.
On the other hand the coupling $c_H$ increases in magnitude with $\tan\beta$ 
for both $\mu>0$ and $\mu<0$, but is positive when $\mu>0$ and negative for 
$\mu<0$. In addition, in the decoupling limit the coupling of the light 
Higgs to down-type quarks is SM-like, and the heavy Higgses couple to 
down-type quarks proportionally to $\tan\beta$. 
The sfermion contribution is dominated by the gauge coupling of the wino-like 
component neutralino to the sfermion and the quarks. We remark that for the 
parameter range under consideration there is destructive interference between 
the amplitude for the Higgs and sfermion-exchange diagrams for $\mu>0$, and 
for $\mu<0$ when~\cite{Crivellin:2015bva}
\begin{equation}
\frac{m_H^2 (1-2/t_\beta)}{m_h^2}< t_\beta,
\end{equation}
provided $M_2\simeq |\mu|$ and $t_\beta\equiv \tan \beta \gg 1$.
For these cases lower values of the sfermion masses reduce the scattering 
cross section.
 
\begin{figure}[t]
  \centering
  \includegraphics[width=.49\textwidth]{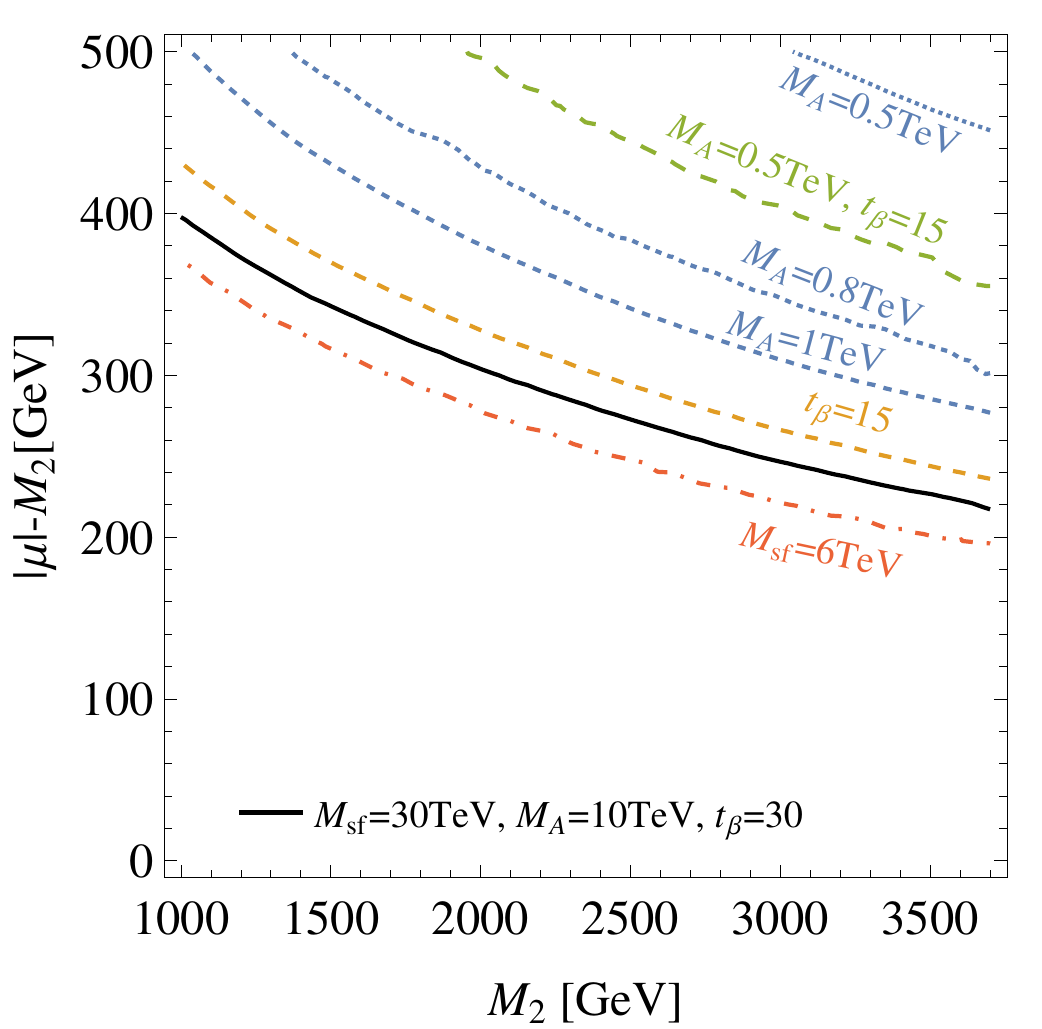}
  \includegraphics[width=.49\textwidth]{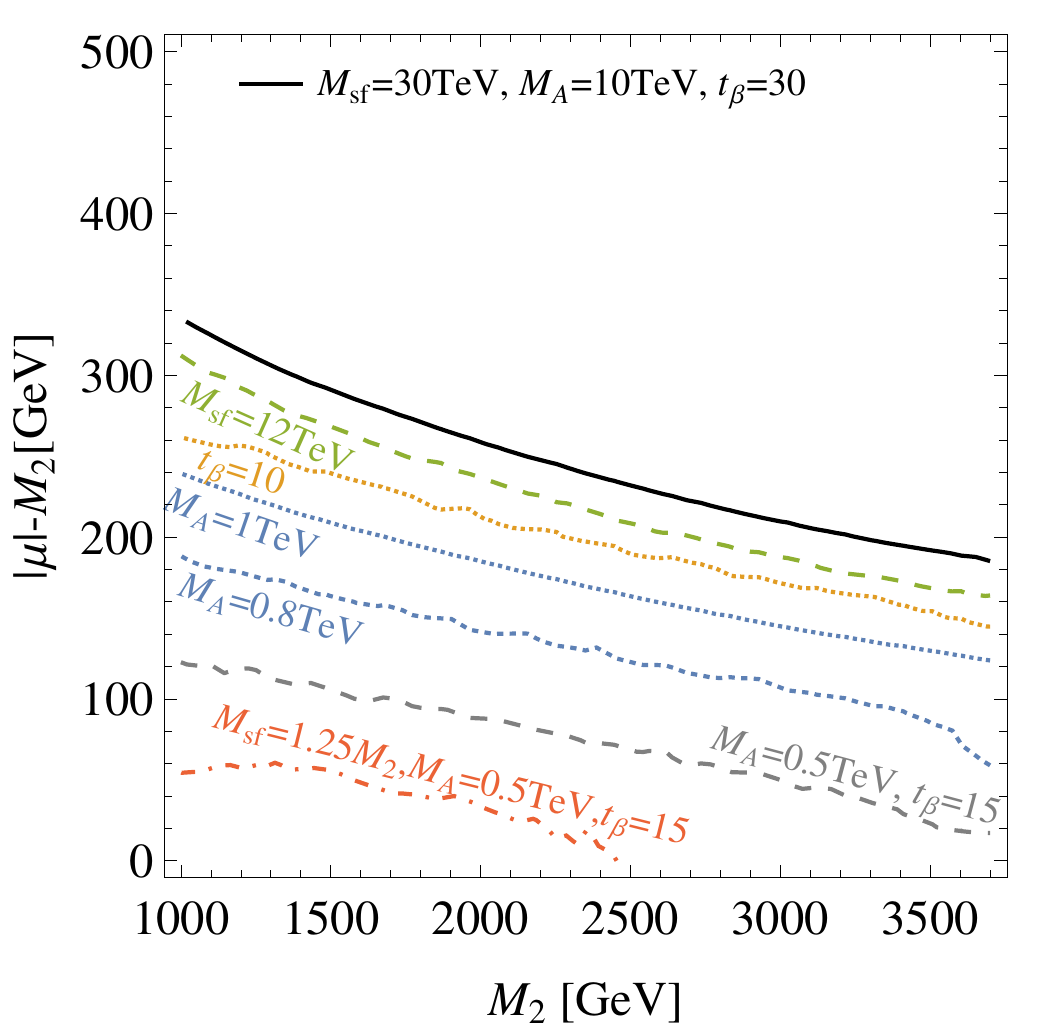}
\caption{Direct detection limits for different choices of the MSSM parameters,
assuming the neutralino is completely responsible for the measured dark 
matter density of the Universe. Where not stated, the parameter choices 
correspond to those for the black line. The area below the lines is excluded. 
The left panel shows the case of $\mu>0$,
while the right of $\mu<0$.\label{fig:DDlimits}}
\end{figure}

In Fig.~\ref{fig:DDlimits} we show the resulting limits from LUX data 
in the $|\mu|-M_2$ vs.\ $M_2$ plane for different choices of  
$t_\beta$, $M_A$, $M_{\rm sf}$, and the sign of $\mu$. The 
above discussion allows us to understand the following trends observed: 
\begin{itemize}
\item On decreasing $t_\beta$ and $M_A$ the direct detection bound becomes 
stronger for positive $\mu$ and weaker for negative $\mu$. Note that for 
$\mu<0$ the cross section decreases, and the bound weakens, due 
to the destructive 
interference between the $h$ and $H$ contributions as the relative sign 
between the couplings $c_h$ and $c_{H}$ changes.
\item The direct detection bound weakens for less decoupled sfermions when 
there is destructive interference between the t-channel Higgs-exchange and 
s-channel sfermion-exchange diagrams. This always occurs for $\mu>0$, while 
for $\mu<0$ one requires small heavy Higgs masses. For instance, for 
$t_\beta=15$ the maximum value of $M_A$ giving destructive interference is 
slightly above 500~GeV, while for $t_\beta=30$ one needs $M_A<700$ GeV. 
\end{itemize}
Since we consider a point in the $|\mu|-M_2$ vs.\ $M_2$ plane to be 
excluded only if it is excluded for {\em any} (allowed) value of the other 
MSSM parameters, this means that the bounds from direct detection 
experiments are weakest for $\mu<0$ in combination with low values of 
$M_{\rm sf}$, $M_A$ and $\tan\beta$, and for $\mu>0$ in combination with 
high values of $M_A$ and $\tan\beta$ but low values of $M_{\rm sf}$.

%%%%%%%%%%%%%%%%%%%%%%%%%%%%%%%%%%%%%%%%%%%%%%%%%%%%%%%%%%%%%%%%%%%%%%%%%%%%%%
%%%%%%%%%%%%%%%%%%%%%%%%%%%%%%%%%%%%%%%%%%%%%%%%%%%%%%%%%%%%%%%%%%%%%%%%%%%%%%

%%%%%%%%%%%%%%%%%%%%%%%%%%%%%%%%%%%%%%%%%%%%%%%%%%%%%%%%%%%%%%%%%%%%%%%%%%%%%%

\section{Results: indirect detection and CMB limits}
\label{sec:results}

In this section we first determine the region of the $|\mu|-M_2$ vs.\ $M_2$ 
plane which satisfies the relic density constraint and is allowed by the  
gamma-ray limits from dwarf spheroidals, the positron limits from AMS-02, 
and the CMB limits.\footnote{For the combined $e^+ + e^-$ flux several 
earlier observations provide data extending to higher energies than the 
AMS-02 experiment, though with much larger uncertainties. We do not include 
these data in our analysis, because for the DM models under consideration, 
the strongest lepton limits arise from energies below about 100 GeV, in 
particular the from observed positron fraction (see Fig.~7 of 
\cite{Hryczuk:2014hpa}).}
We also determine the regions preferred by fits to 
AMS-02 antiproton results. Over a large part of the considered $|\mu|-M_2$ 
vs.\ $M_2$ plane, the observed relic density can be obtained for some value 
of the sfermion masses and other MSSM parameters. For the remaining region of 
the plane, where the relic density constraint is not fulfilled for thermally 
produced neutralino dark matter, we consider both, the case where the dark 
matter density is that observed throughout the plane, in which case it 
cannot be produced thermally, and the case where it is always thermally 
produced, for which the neutralino relic density does not always agree with 
that observed, and the limits must be rescaled for each point in the plane 
by the relic density calculated accordingly. That the neutralino dark matter 
is not thermally produced, or that it only constitutes a part of the total 
dark matter density are both viable possibilities.

We then consider various slices through this plane for fixed values of 
$|\mu|-M_2$, and show the calculated present-day annihilation cross section as 
a function of $M_2\sim m_{\chi_1^0}$ together with the same limits and 
preferred regions as above, both for the case that the limits are and are 
not rescaled according to the thermal relic density.

%%%%%%%%%%%%%%%%%%%%%%%%%%%%%%%%%%%%%%%%%%%%%%%%%%%%%%%%%%%%%%%%%%%%%%%%%%%%%%

\subsection{Limits on mixed-wino DM}
\label{sec:limits-obsrd}

In this section we present our results on the limits from indirect searches 
for wino-like DM in the MSSM, assuming the relic density is as observed. 
That is, for most parameter points the DM must be produced non-thermally or 
an additional mechanism for late entropy production is at play. We show each 
of the considered indirect search channels separately in the $|\mu|-M_2$ 
vs.~$M_2$ plane (including both $\mu>0$ and $\mu<0$), superimposing on this 
the contours of the correct relic density for three choices of the sfermion 
mass. Note that while the indirect detection limits are calculated for 
$M_{\rm sf}=8$ TeV, the effect of the choice of sfermion mass on them is 
minimal, and therefore we display only the relic density contours 
for additional values of $M_{\rm sf}$.

%%%%%%%%%%%%%%%%%%%%%%%%%%%%%%%%%%%%%%%%%%%%%%%%%%%%%%%%%%%%%%%%%%%%%%%%%%%%%%
\begin{figure}[t!]
  \centering
  \includegraphics[width=.49\textwidth]{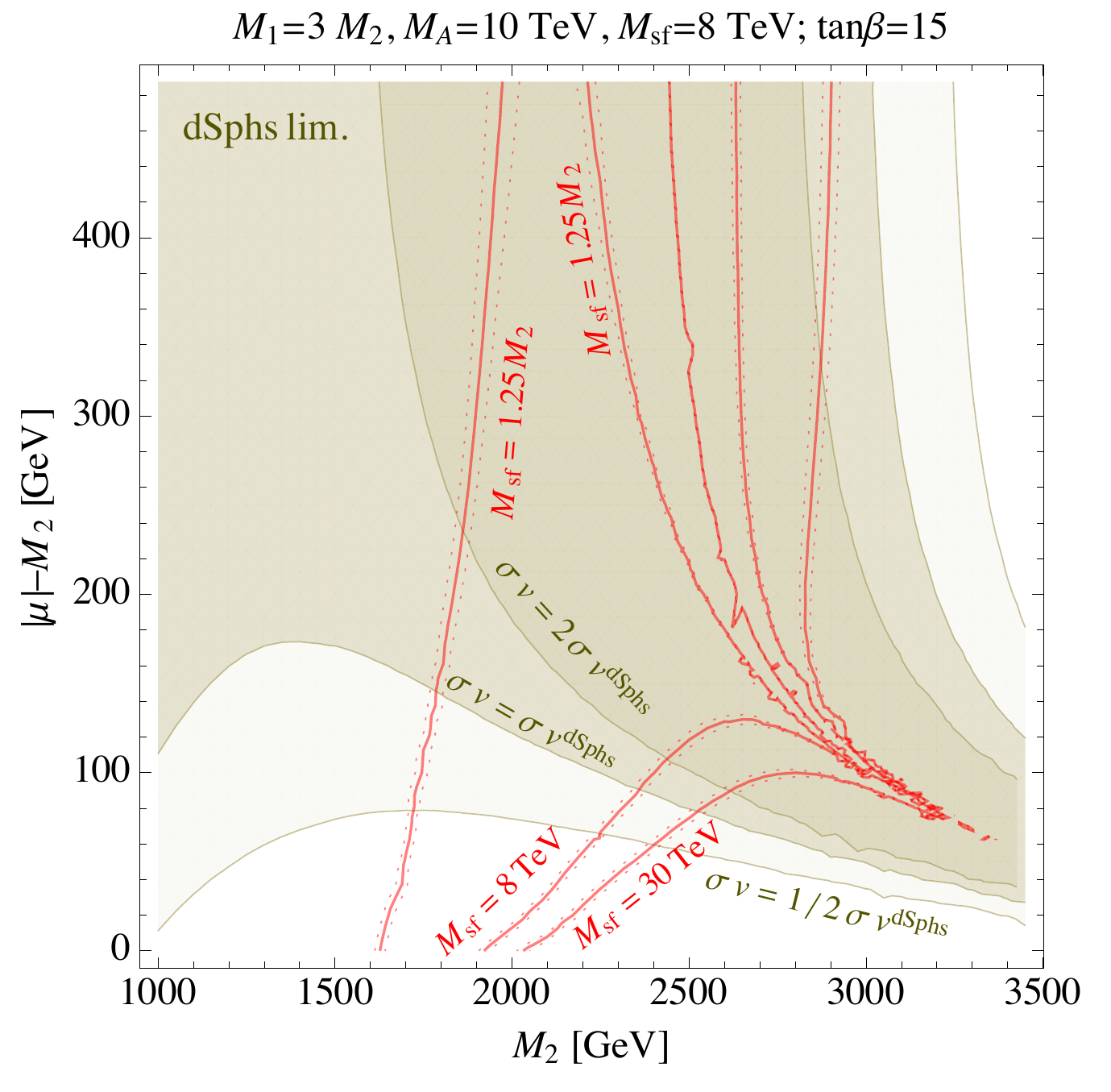}
  \includegraphics[width=.49\textwidth]{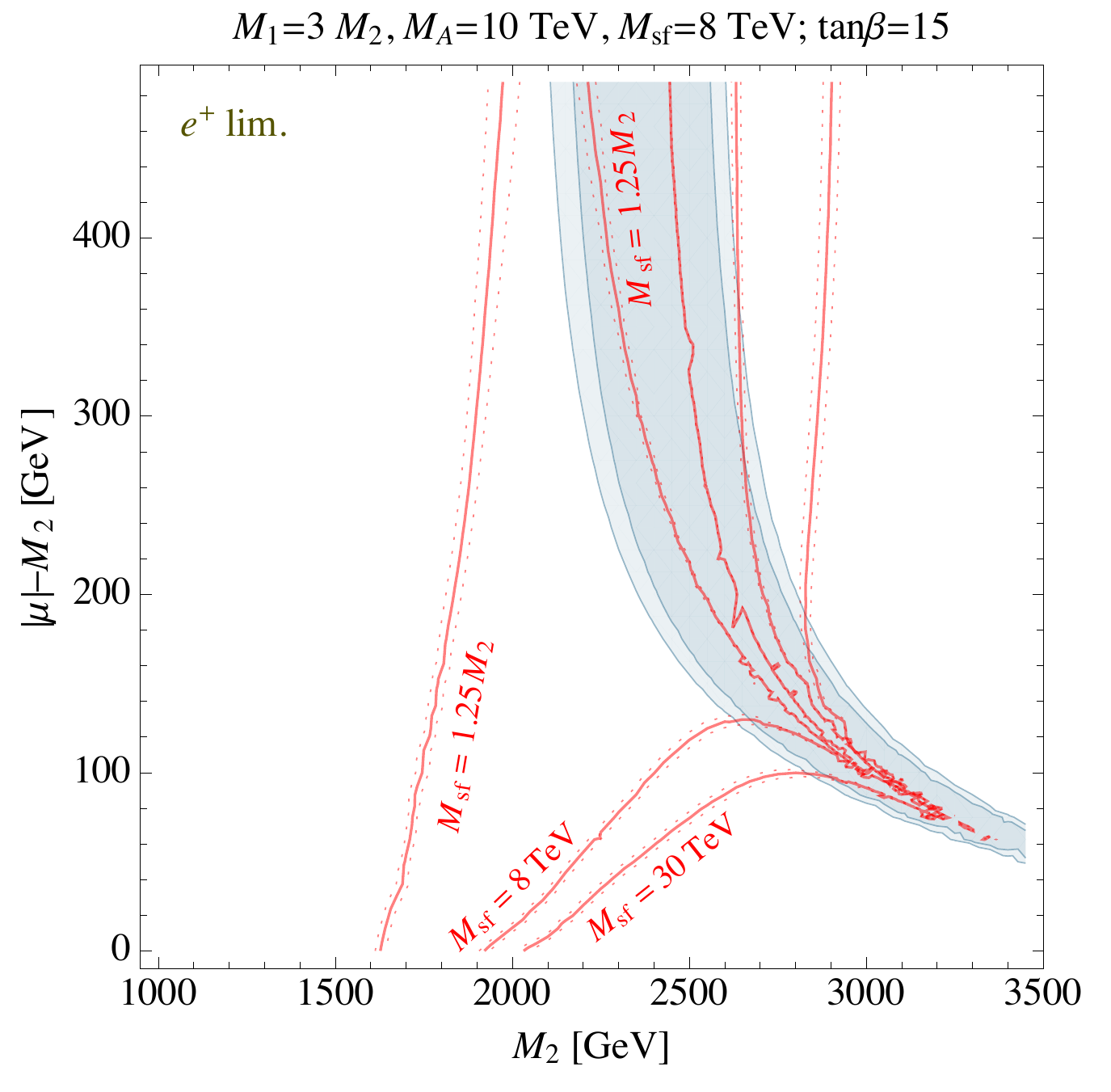}\\
   \includegraphics[width=.49\textwidth]{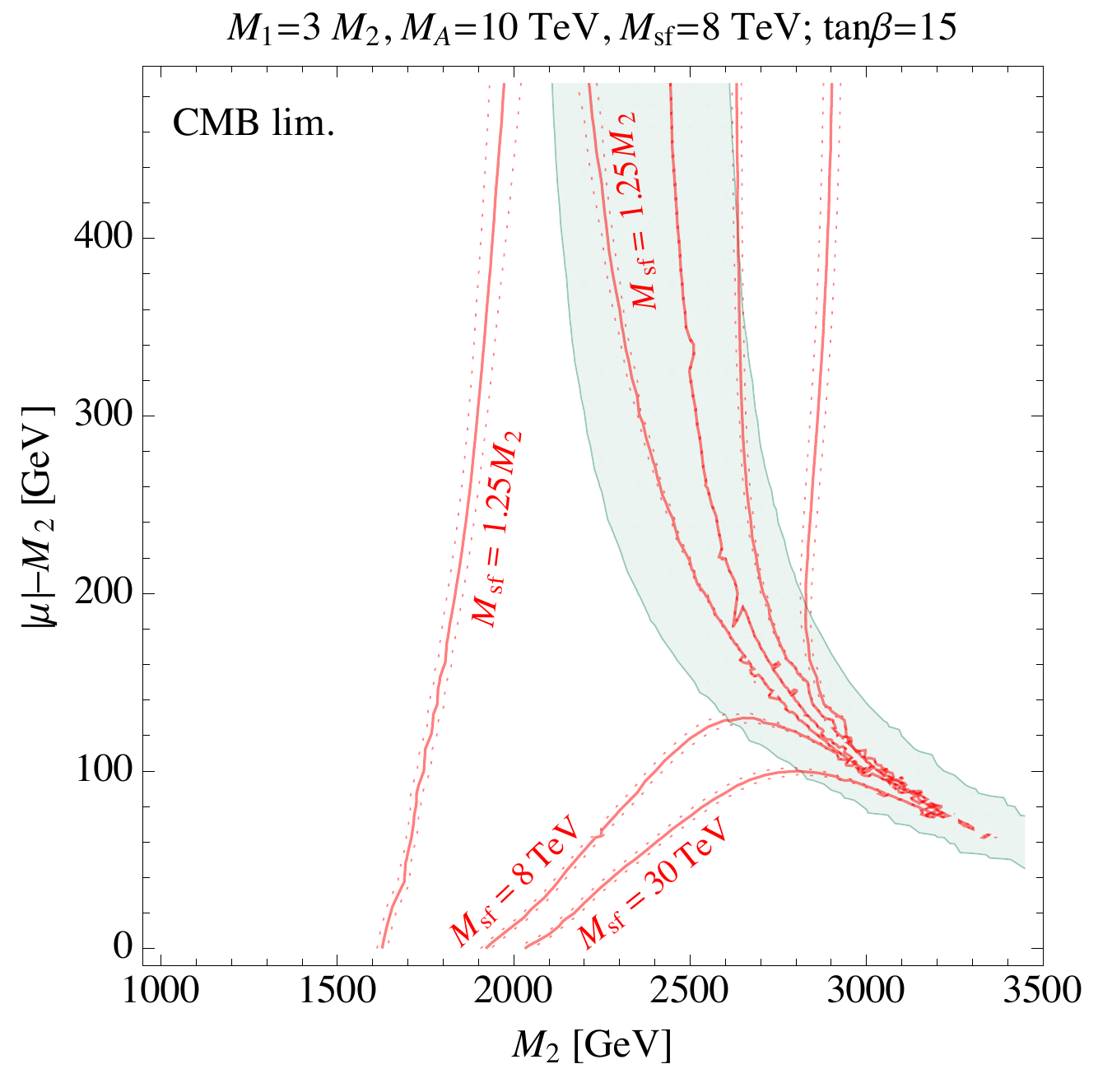}
  \includegraphics[width=.49\textwidth]{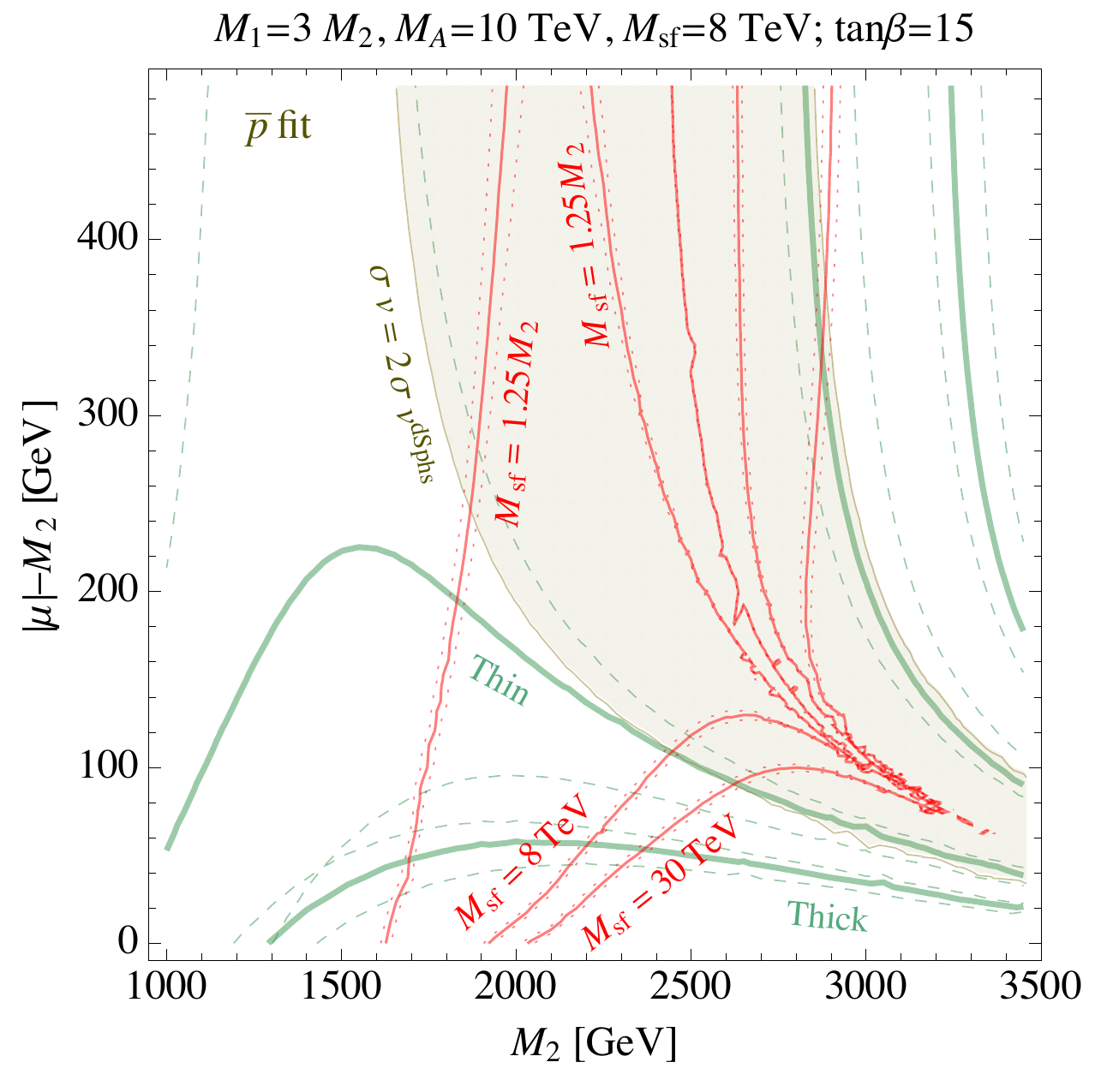}
\caption{Results in the $M_2$ vs.\ $|\mu|-M_2$ plane. Left: limits from dSphs 
(upper) and the CMB (lower). The shaded regions are excluded, different 
shadings correspond to the DM profile uncertainty. Right: the region excluded 
by AMS-02 leptons (upper), and the best fit contours for antiprotons (lower), 
where the green solid lines show the Thin and Thick propagation models, 
while the dotted lines around them denote the 1$\sigma$ confidence 
intervals. Contours where the observed relic density is obtained for the 
indicated value of the sfermion mass are overlaid.
\label{fig:higgsino}}
\end{figure}
%%%%%%%%%%%%%%%%%%%%%%%%%%%%%%%%%%%%%%%%%%%%%%%%%%%%%%%%%%%%%%%%%%%%%%%%%%%%%%

In Fig.~\ref{fig:higgsino} we show the exclusions from dSphs, $e^+$, and the 
CMB separately in the $|\mu|-M_2$ vs.\ $M_2$ plane. For the positrons we show 
two limits, obtained on assuming the Thin and Thick propagation models 
described in Section~\ref{sec:Bckgr}. We see that the most relevant 
exclusions come from the diffuse gamma-ray searches from dSphs. Here we show 
three lines corresponding  to the limit on the cross section assuming the 
Navarro-Frenk-White profile in dSphs, and rescaling this limit 
up and down by a factor 2. This is done in order to estimate the effect 
of the uncertainty in the $J$-factors. For instance, the recent 
reassessment \cite{Ullio:2016kvy} of the $J$-factor for Ursa Minor inferred 
from observational data suggests 2 to 4 times smaller limits than those 
commonly quoted. In order to provide conservative bounds, we adopt the 
weakest of the three as the reference limit. We then compare (lower right 
plot) this weakest limit from dSphs to the preferred region obtained on 
fitting to the AMS-02 antiproton results, showing the results for both 
Thin and Thick propagation models.\footnote{The actual analysis was finalized 
before the recent antiproton results were published \cite{Aguilar:2016kjl} 
and hence was based on earlier data presented by the AMS 
collaboration~\cite{AMSdays1}. This is expected to have a small effect on the 
antiproton fit presented in this work, with no significant consequences for 
the overall results.}

We find that there are parts of the mixed wino-Higgsino and dominantly 
wino neutralino parameter space both below and above the Sommerfeld resonance 
region, where the relic density is as observed and which are compatible with 
the non-observation of dark matter signals in indirect searches. 
In the lower right plot of Fig.~\ref{fig:higgsino} we see that 
these further overlap with the regions preferred by fits to the antiproton 
results. In the smaller region above the resonance, this overlap occurs when 
the sfermions are decoupled, and hence corresponds to an almost pure-wino 
case, whereas below the resonance the overlap region is spanned by 
varying the sfermion masses from $1.25 M_2$ to being decoupled. The latter region requires 
substantial Higgsino-mixing of the wino, and extends from $M_2=1.7$ TeV to 
about 2.5 TeV, thus allowing dominantly-wino dark matter in a significant 
mass range.

\begin{figure}[t]
\centering
\includegraphics[width=.49\textwidth]{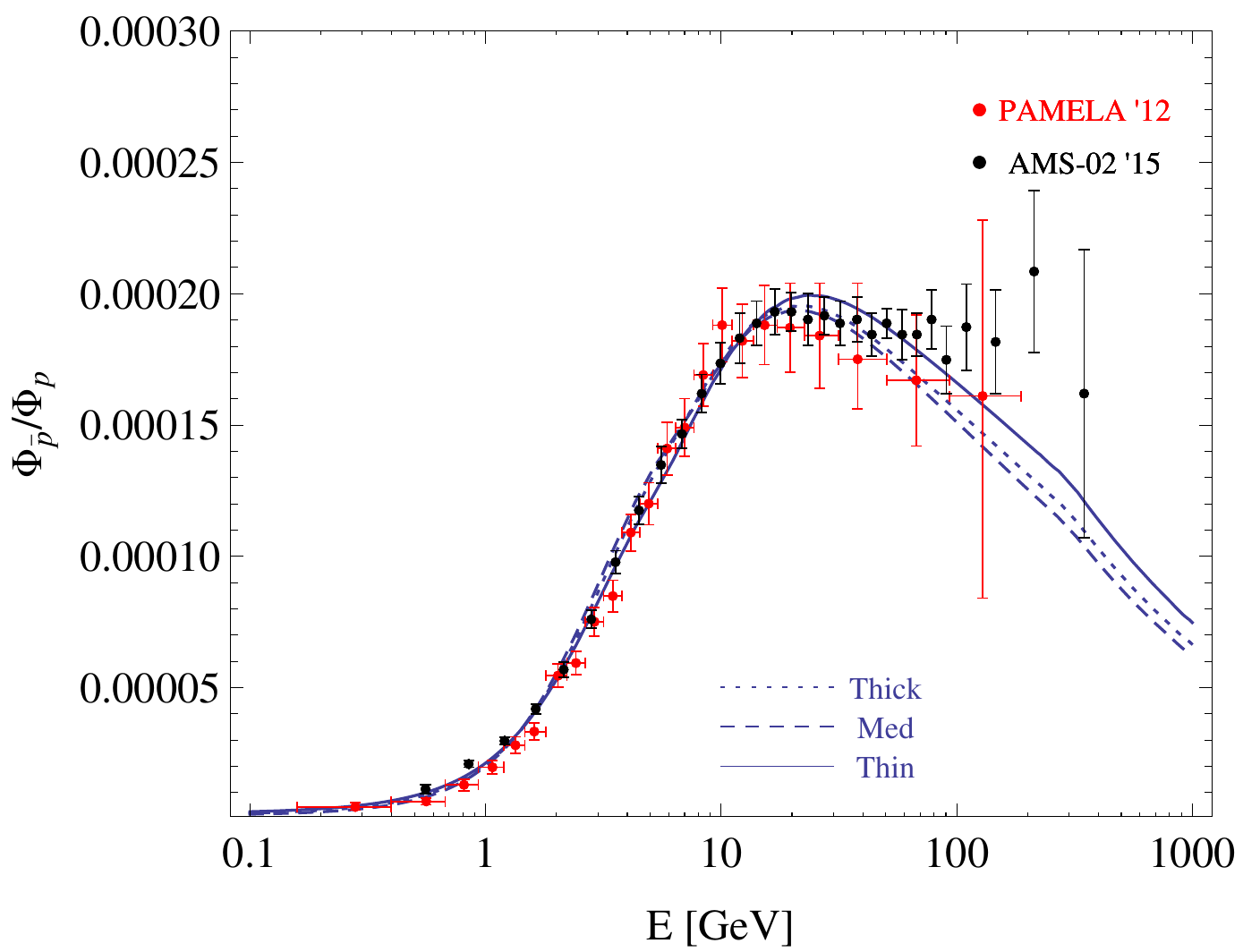}
\includegraphics[width=.49\textwidth]{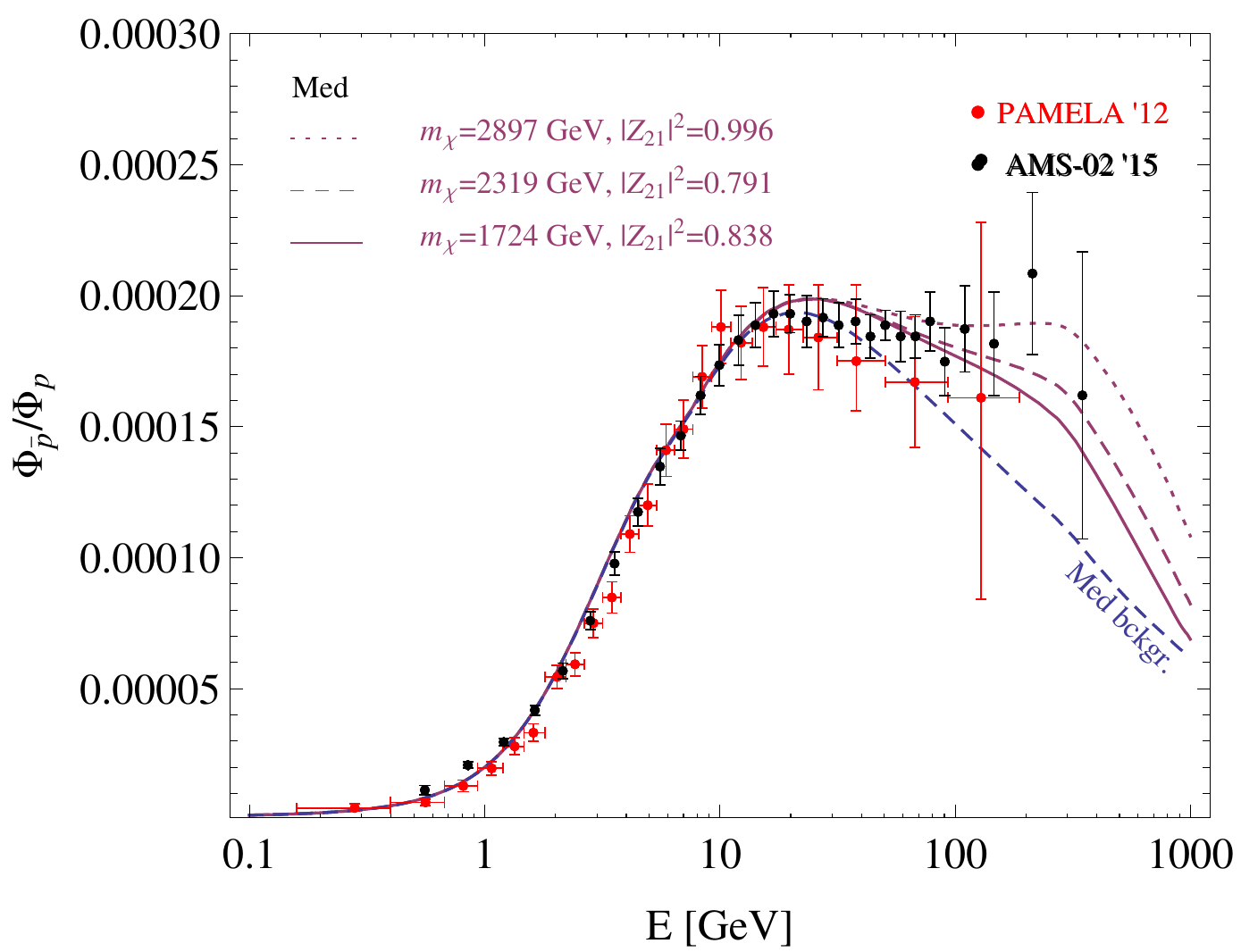}
\caption{The antiproton-to-proton ratio: background propagation models 
(left) and comparison of three DM models with relic density within the 
observational range and assuming the ``Med'' propagation (right). The shown 
data is from AMS-02 \cite{AMSdays1} and PAMELA~\cite{Adriani:2012paa}. 
\label{fig:antip}}
\end{figure}

Let us comment on the improvement of the fit to the antiproton measurements 
found for some choices of the parameters. In Fig.~\ref{fig:antip} we show 
examples of antiproton-to-proton ratio fits to the data from the background 
models (left) and including the DM component (right). Although the 
propagation and antiproton production uncertainties can easily resolve the 
apparent discrepancy of the background models vs.\ the observed 
data \cite{Kappl:2015bqa,Evoli:2015vaa,Giesen:2015ufa}, it is nevertheless 
interesting to observe that the spectral shape of the DM component matches 
the observed data for viable mixed-wino dark matter 
particles.

%%%%%%%%%%%%%%%%%%%%%%%%%%%%%%%%%%%%%%%%%%%%%%%%%%%%%%%%%%%%%%%%%%%%%%%%%%%%%%
%%%%%%%%%%%%%%%%%%%%%%%%%%%%%%%%%%%%%%%%%%%%%%%%%%%%%%%%%%%%%%%%%%%%%%%%%%%%%%

%%%%%%%%%%%%%%%%%%%%%%%%%%%%%%%%%%%%%%%%%%%%%%%%%%%%%%%%%%%%%%%%%%%%%%%%%%%%%%

\subsection{Indirect search constraints on the MSSM 
parameter space}
\label{sec:limits-thermrd}

In this section we present our results for the limits from indirect searches 
on wino-like DM, assuming the relic density is always thermally produced. In 
other words, for the standard cosmological model, these constitute the limits 
on the parameter space of the MSSM, since even if the neutralino does not 
account for all of the  
dark matter, its thermal population can give large enough signals to be seen 
in indirect searches. In this case a parameter-space point is excluded, if 
\begin{equation}
(\sigma v)_0 \big|_{\rm th}> \left(\frac{\Omega h^2|_{\rm obs}}
{\Omega h^2|_{\rm thermal}}\right)^2 (\sigma v)_0 \big|_{\rm exp\,lim}\,
\label{eq:rescaling}
\end{equation}
where $(\sigma v)_0 \big|_{\rm th}$ is the theoretically predicted 
present-day cross section and $(\sigma v)_0 \big|_{\rm exp\,lim}$ the limit quoted 
by the experiment. This is because the 
results presented by the experiments assume the DM particle to account for the 
entire observed relic density.
Therefore if one wishes to calculate the limits for dark matter candidates 
which only account for a fraction of the relic density, one needs to rescale 
the bounds by the square of the ratio of observed relic density 
$\Omega h^2|_{\rm obs}$ to the thermal relic density 
$\Omega h^2|_{\rm thermal}$. Viewed from another perspective, the results 
below constitute astrophysical limits on a part of the MSSM parameter space,
which is currently inaccessible to collider experiments, with the only 
assumption that there was no significant entropy production in the early 
Universe after the DM freeze-out.

%%%%%%%%%%%%%%%%%%%%%%%%%%%%%%%%%%%%%%%%%%%%%%%%%%%%%%%%%%%%%%%%%%%%%%%%%%%%%%
\begin{figure}[t!]
  \centering
  \includegraphics[width=.49\textwidth]{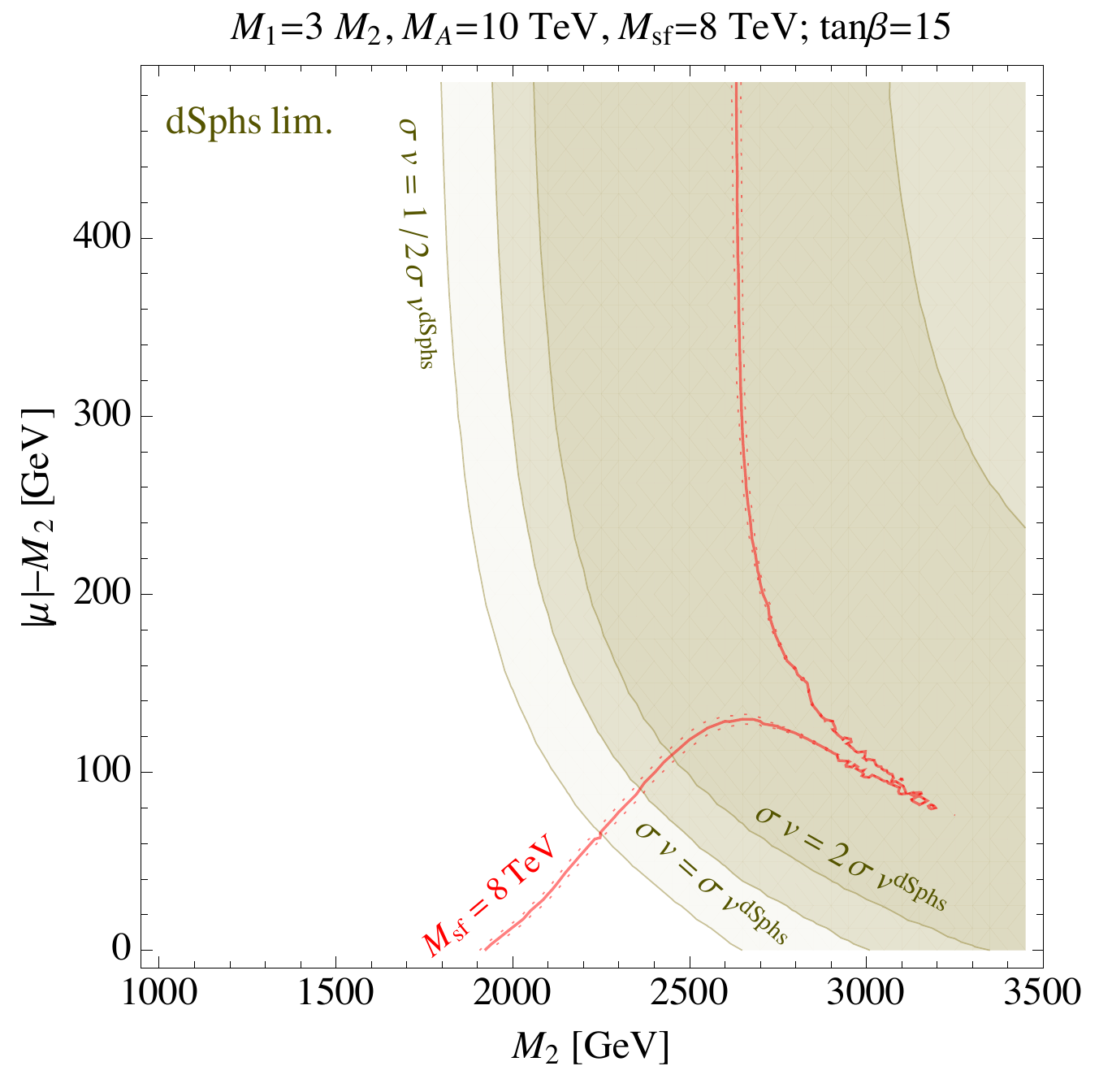}
  \includegraphics[width=.49\textwidth]{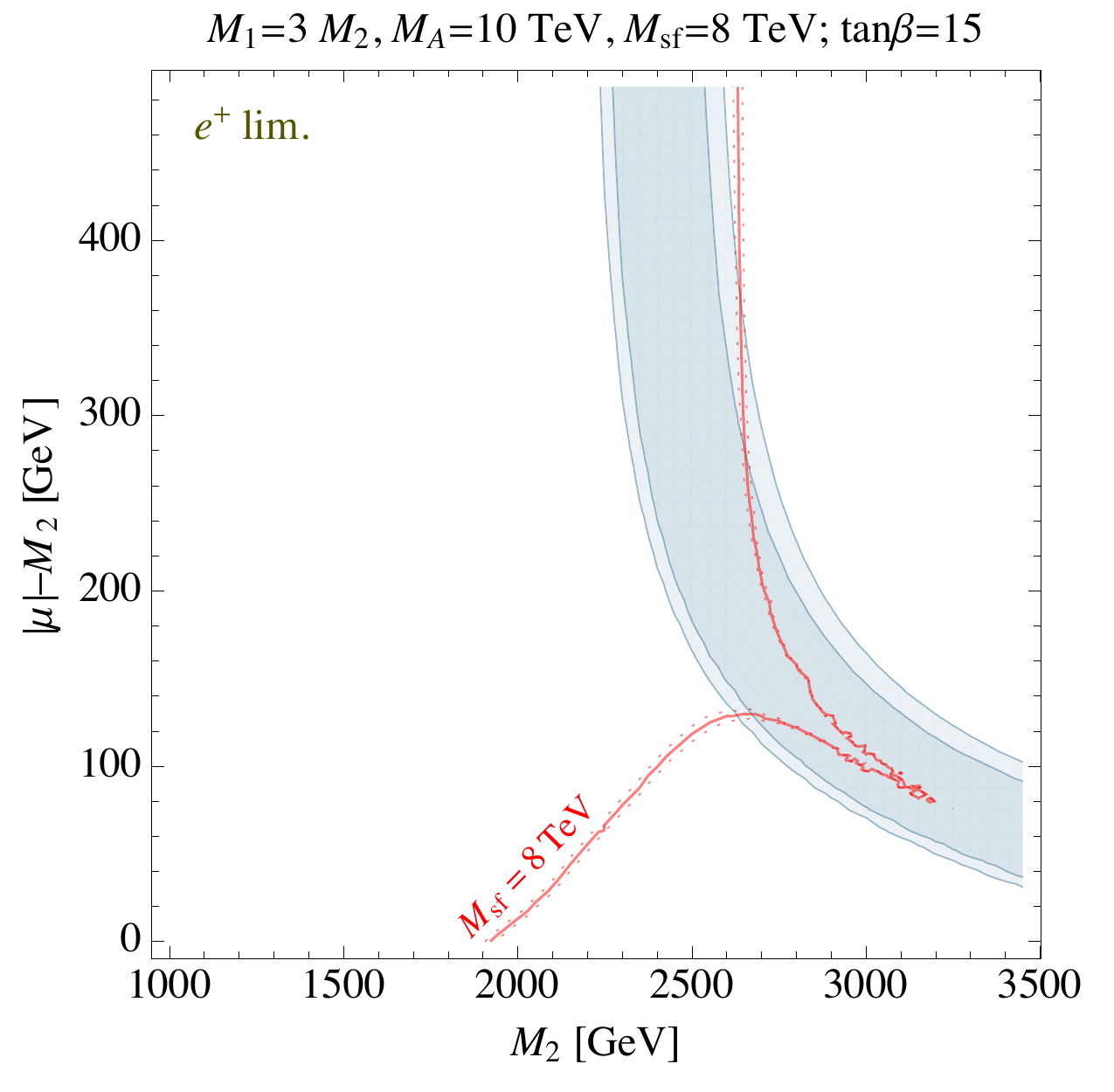}\\
  \includegraphics[width=.49\textwidth]{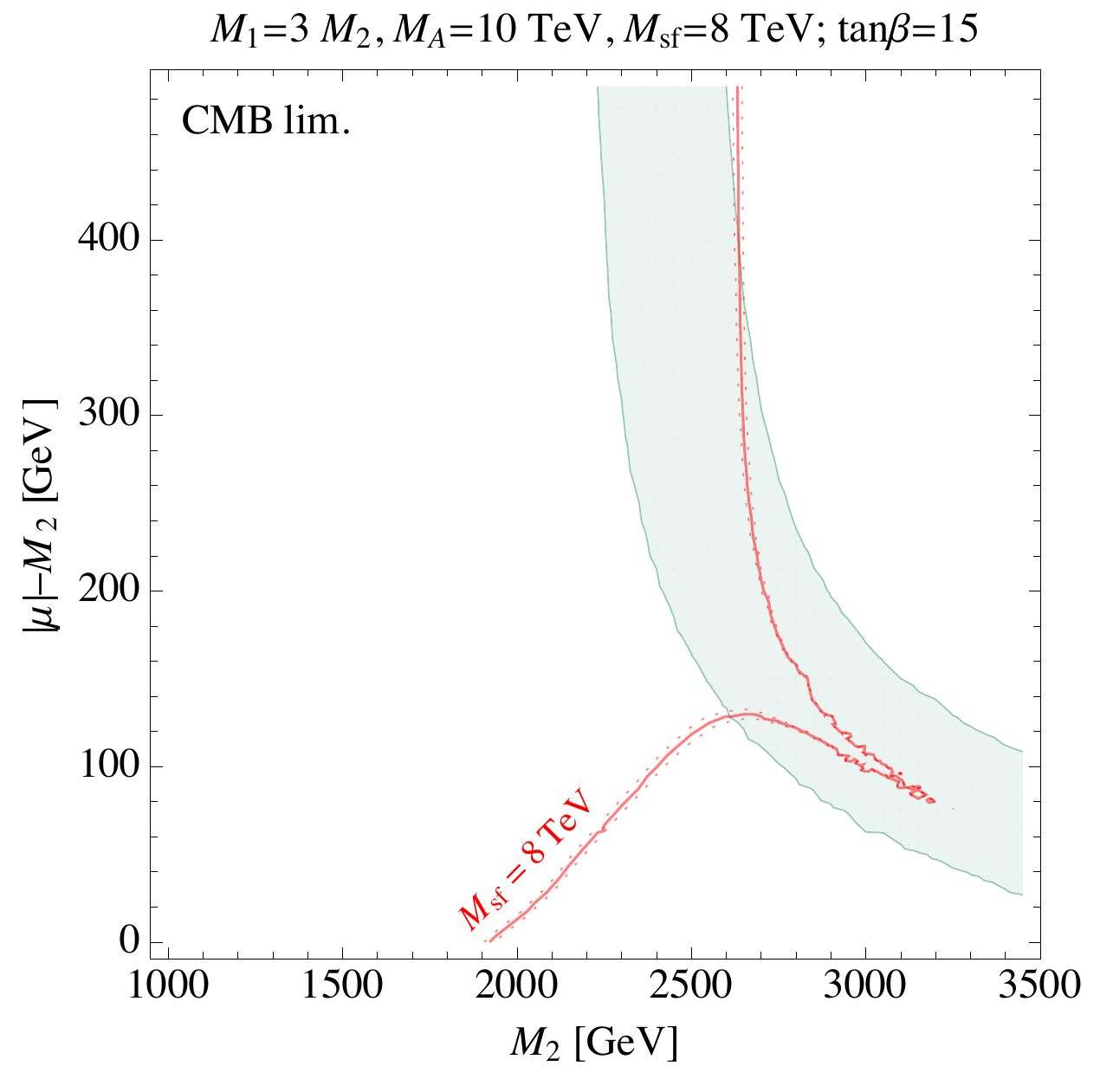}
  \includegraphics[width=.49\textwidth]{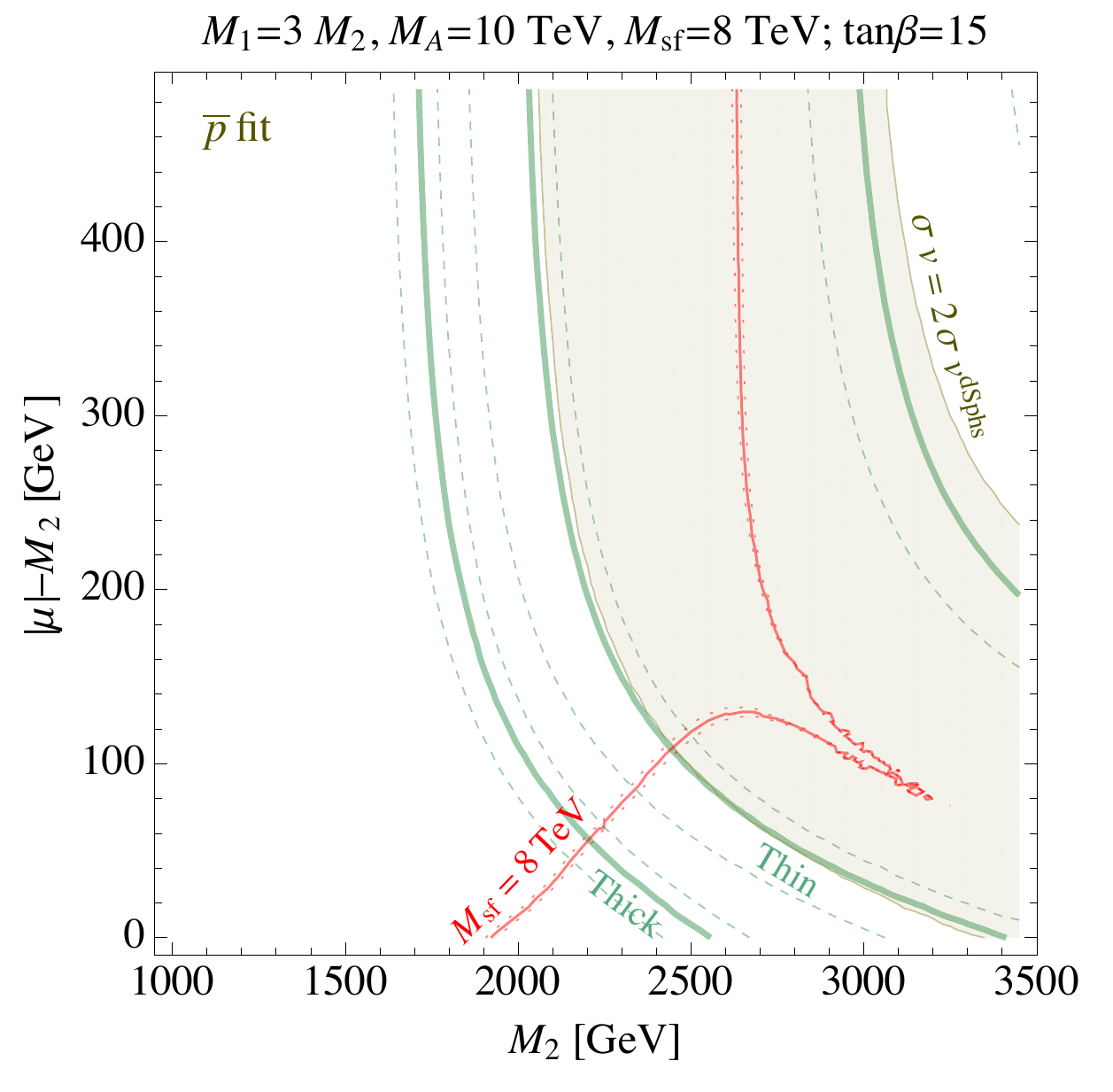}
\caption{Results in the $M_2$ vs $|\mu|-M_2$ plane for the case where the 
limits are rescaled according to the thermal relic density for a given point 
in the plane. Details are as in Fig.~\ref{fig:higgsino}.
\label{fig:higgsino_rescaled}}
\end{figure}
%%%%%%%%%%%%%%%%%%%%%%%%%%%%%%%%%%%%%%%%%%%%%%%%%%%%%%%%%%%%%%%%%%%%%%%%%%%%%%

In Fig.~\ref{fig:higgsino_rescaled}, as in the previous subsection, we show 
the exclusions from dSphs, $e^+$, and the CMB individually in the $|\mu|-M_2$ 
vs.\ $M_2$ plane. The limits are calculated as for Fig.~\ref{fig:higgsino}.
We then compare the weakest limit from dSphs to the preferred region 
obtained on fitting to the AMS-02 antiproton results, where we show the
results for both Thin and Thick propagation models.
Again we find that parameter regions exist where the relic density is correct 
and which are not excluded by indirect searches. The marked difference 
between the previous and present results is that in 
Fig.~\ref{fig:higgsino_rescaled} the region of the plots for lower $M_2$ is 
not constrained by the indirect searches, because in this region the 
thermal relic density is well below the measured value and therefore the 
searches for relic neutralinos are much less sensitive. 
In the bottom lower plot of 
Fig.~\ref{fig:higgsino_rescaled} we see that the unconstrained regions 
overlap with the regions preferred by fits to the antiproton results. 
While the limits themselves do not depend on the sfermion mass, the 
thermal relic density does, and therefore the rescaling of the limits 
via~\eqref{eq:rescaling} induces a dependence on the sfermion mass. Therefore 
the intersection of the lines of correct relic density for 
$M_{\rm sf}\neq 8$ TeV with the preferred region from antiproton searches 
is not meaningful, and we do not show them in the plots.

%%%%%%%%%%%%%%%%%%%%%%%%%%%%%%%%%%%%%%%%%%%%%%%%%%%%%%%%%%%%%%%%%%%%%%%%%%%%%%

\subsection{\boldmath
Limits on the present-day cross section for fixed $|\mu|-M_2$}
\label{sec:sectionplots}

%%%%%%%%%%%%%%%%%%%%%%%%%%%%%%%%%%%%%%%%%%%%%%%%%%%%%%%%%%%%%%%%%%%%%%%%%%%%%%
\begin{figure}[t]
  \centering
  \includegraphics[width=.72\textwidth]{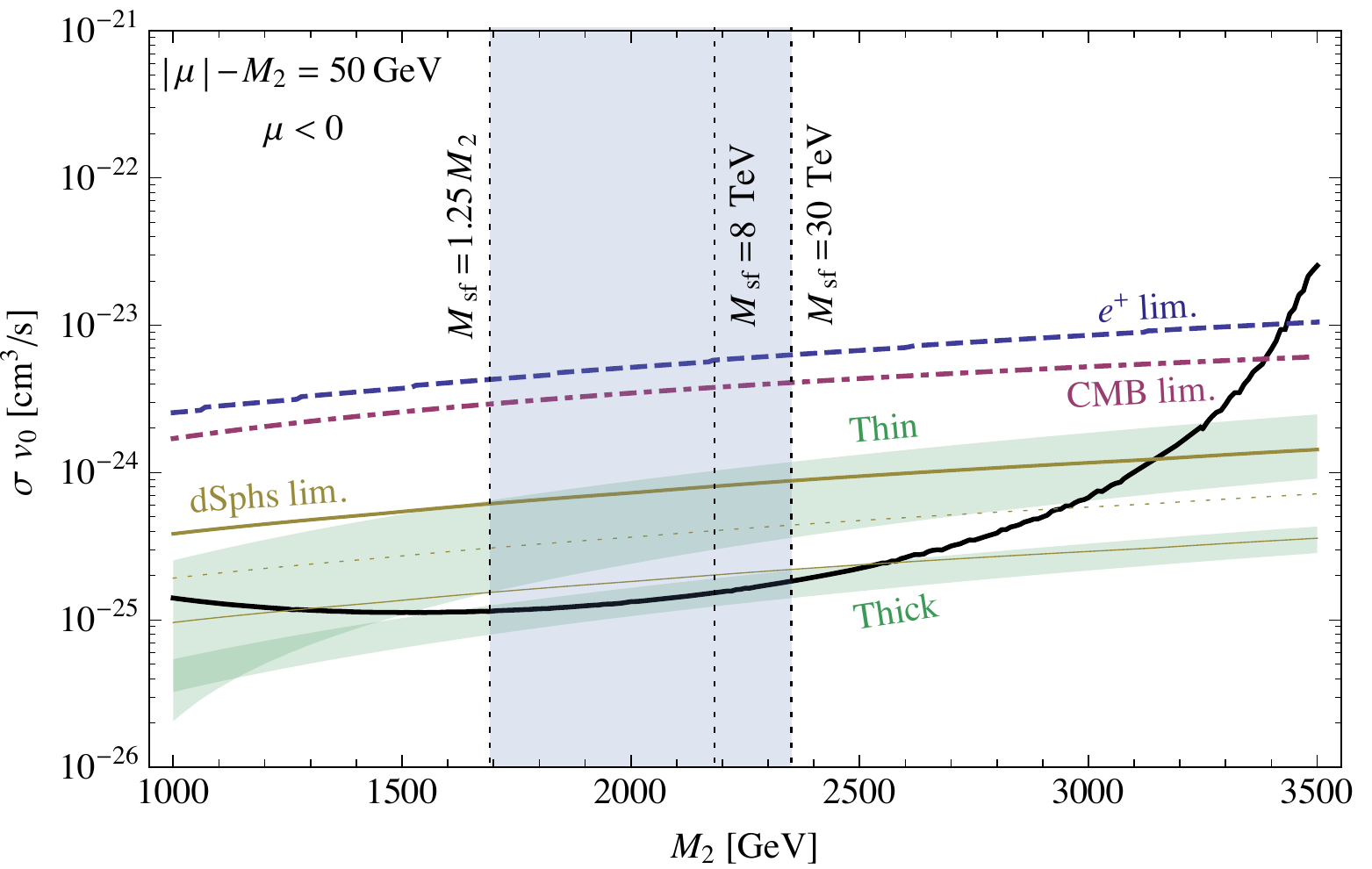} \\
  \includegraphics[width=.72\textwidth]{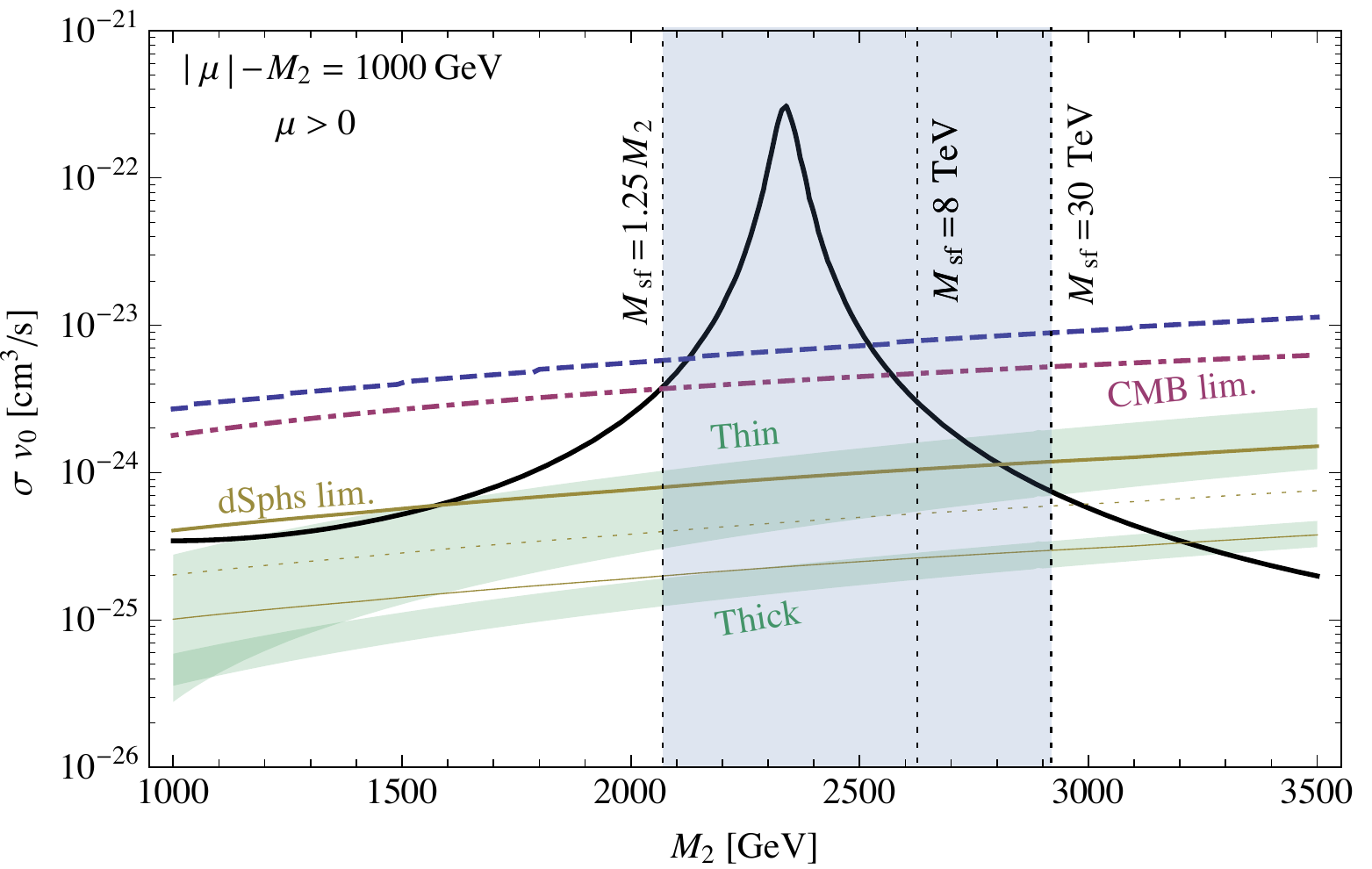}
\caption{The predicted present-day annihilation cross section $(\sigma v)_0$ 
(black) is shown as a function of $M_2\sim m_{\chi_1^0}$ for the Higgsino 
admixture $|\mu|-M_2$ as indicated. This is compared with exclusion limits 
from dSphs (brown), positrons (blue dashed) and the CMB (magenta dot-dashed), 
along with the preferred regions from antiproton searches (pale green) 
adopting the Thin and Thick models. We also show the dSphs exclusion limits 
multiplied and divided by 2 (brown), the weaker of which is the thicker line. 
The observed relic density is assumed. The blue shaded region indicates 
where the relic density can correspond to the observed value by changing 
$M_{\rm sf}$.
\label{fig:SectionPlots}}
\end{figure}

\begin{figure}[t]
  \centering
   \includegraphics[width=.72\textwidth]{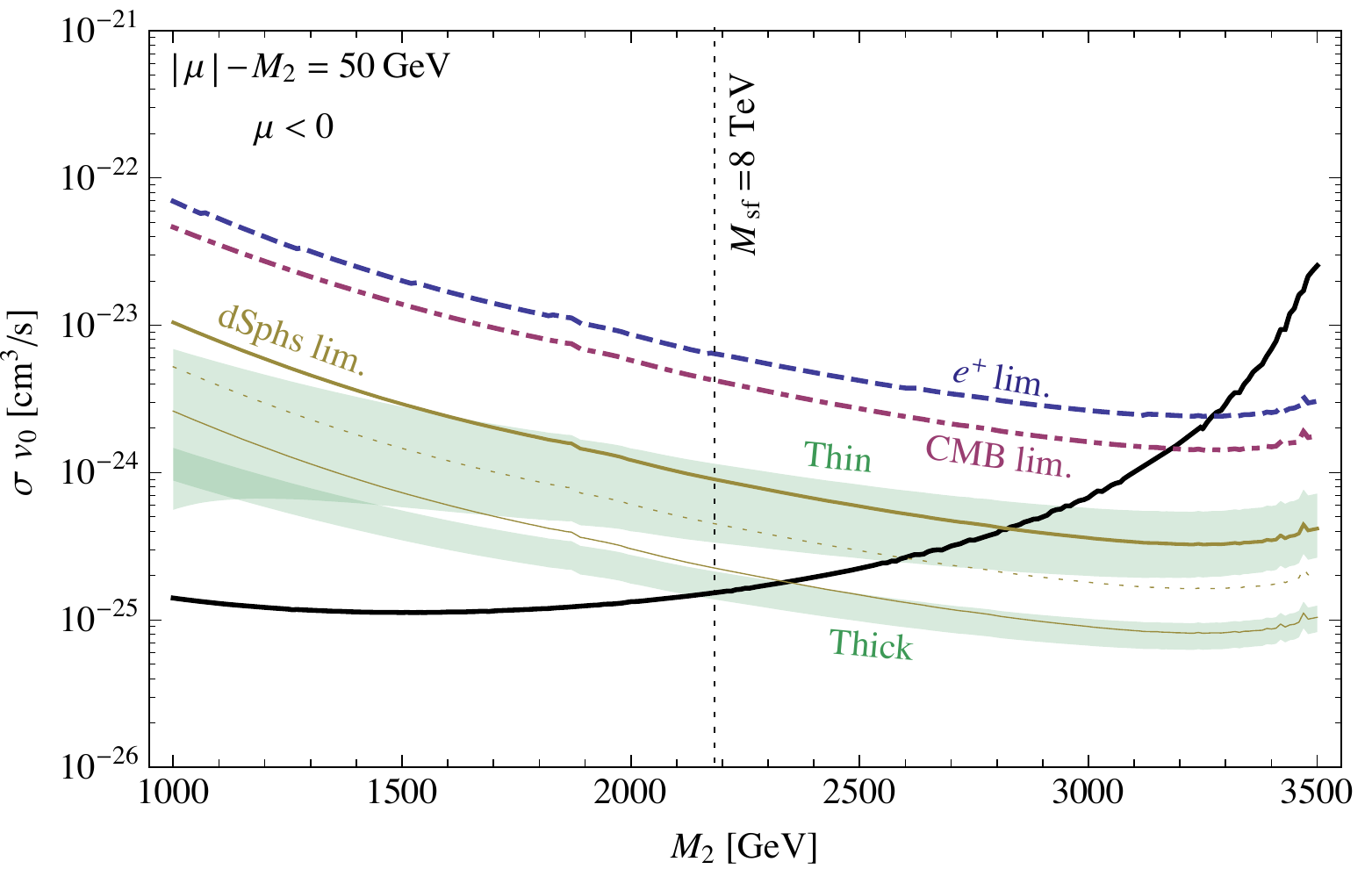}\\
   \includegraphics[width=.72\textwidth]{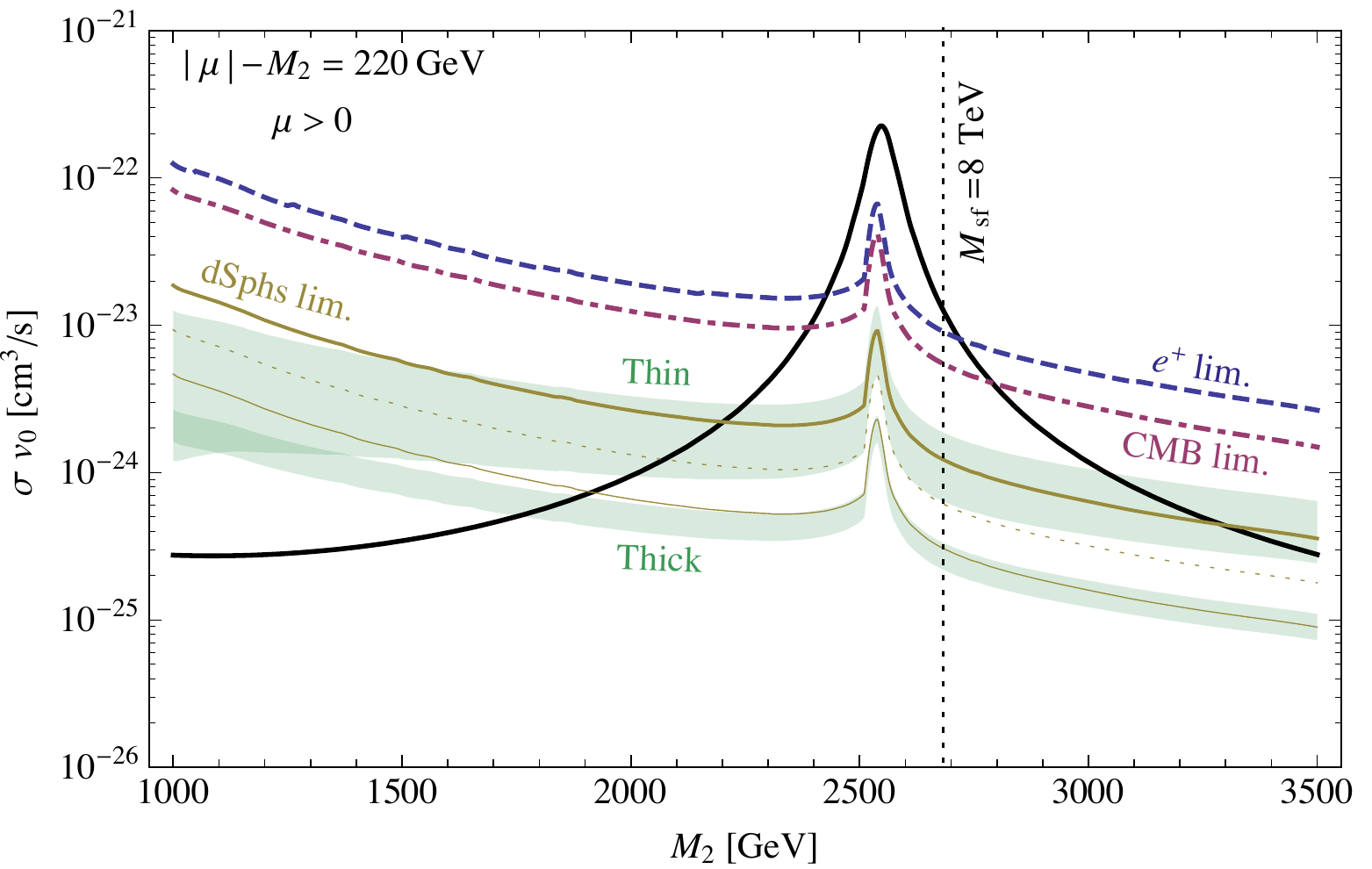}
\caption{As in Fig.~\ref{fig:SectionPlots}, but the thermal relic density 
is assumed and the limits are rescaled according to \eqref{eq:rescaling}. 
Note the different value of $|\mu|-M_2$ in the lower plot compared to 
the previous figure. 
The black-dashed vertical line indicates where the relic density is equal 
to that observed for the 
sfermion mass value $M_{\rm sf}=8$ TeV.
\label{fig:SectionPlots2}}
\end{figure}
%%%%%%%%%%%%%%%%%%%%%%%%%%%%%%%%%%%%%%%%%%%%%%%%%%%%%%%%%%%%%%%%%%%%%%%%%%%%%%

In order to understand how the limits and the present-day annihilation 
cross section depend on the mass of the DM candidate, we take slices of 
the $|\mu|-M_2$ vs.  
$M_2$ plane for fixed values of $|\mu|-M_2$, and plot $(\sigma v)_0$ (black) 
as a function of $M_2$, which is approximately equal to the LSP mass 
$m_{\chi_1^0}$ in the range shown in Figs.~\ref{fig:SectionPlots} and 
\ref{fig:SectionPlots2}. As in Figs.~\ref{fig:higgsino} and 
\ref{fig:higgsino_rescaled} we show the limits from dSphs (brown), positrons 
(blue dashed) and the CMB (magenta dot-dashed), along with the preferred 
regions from antiproton searches (pale green) adopting the Thin and Thick 
propagation models. We consider three choices of $\mu-M_2$: a very mixed 
neutralino LSP, $|\mu|-M_2=50$~GeV where $\mu$ is negative, 
a mixed case $|\mu|-M_2=220$~GeV where $\mu$ is positive, and
an almost pure-wino scenario, $|\mu|-M_2=1000$~GeV.
The blue shaded region indicates where the relic density can correspond to 
the observed value by changing $M_{\rm sf}$. 

For Fig.~\ref{fig:SectionPlots} we adopt the unrescaled limit, that is, 
two sections of Fig.~\ref{fig:higgsino}. In the case of the very mixed 
wino-Higgsino shown in the upper panel there is a wide range of neutralino 
masses for which the black curve lies below the conservative dSphs limit 
and simultaneously within the range of correct relic density spanned by 
the variation of the sfermion mass. This is different for the almost 
pure-wino scenario shown in the lower panel, where only a small mass region 
survives the requirement that the conservative dSphs limit is respected and 
the observed relic density is predicted. Moreover, in this mass region 
the sfermions 
must be almost decoupled. Fig.~\ref{fig:SectionPlots2} shows two cases 
of mixed wino-Higgsino dark matter, which exhibit similar features, but 
now for the case of assumed thermal relic density, such that the limits 
are rescaled.

It is evident from both figures that for lower 
values of $|\mu|-M_2$, larger regions in $M_2$ can provide both the correct 
relic density and present-day cross section below the 
dSphs bounds. We also see that while the correct relic density can be 
attained at the Sommerfeld resonance, the mass regions compatible with 
indirect search constraints typically lie below the Sommerfeld resonance, 
as was evident from Figs.~\ref{fig:higgsino} and~\ref{fig:higgsino_rescaled} .

%%%%%%%%%%%%%%%%%%%%%%%%%%%%%%%%%%%%%%%%%%%%%%%%%%%%%%%%%%%%%%%%%%%%%%%%%%%%%%

\section{Results: including direct detection limits}
\label{sec:maximalplot}

We have seen in the previous section that there is a sizeable mixed 
wino-Higgsino MSSM parameter space where the lightest neutralino has 
the correct relic abundance and evades indirect detection constraints. 
A significant Higgsino fraction might, however, be in conflict with 
the absence of a direct detection signal. In this section we therefore 
combine the exclusion limits from indirect searches 
studied in the previous section with those coming from the latest LUX results 
for direct detection, in order to determine the allowed mixed 
wino-Higgsino or dominantly-wino dark matter parameter space. To this end we 
first determine the maximal region in this space that passes relic density 
and indirect detection limits in the following way.  For a given $|\mu|-M_2$ 
we identify two points in $M_A$, $M_{\rm sf}$ and $\tan\beta$ within the 
considered parameter ranges, i.e.~$M_A\in\{0.5\,\mathrm{TeV}, 10\,
\mathrm{TeV}\}$, $M_{\rm sf}\in\{1.25 M_2, 30\,\mathrm{TeV}\}$ and 
$\tan\beta\in\{5,30\}$,\footnote{Moving the lower limit $M_A=500$ GeV to 
800 GeV would result in a barely noticeable change to the boundaries marked 
by p2.} corresponding to maximal and minimal values of $M_2$, for which the 
relic density matches the observed value. Two distinct areas of parameter 
space arise: the first is larger and corresponds to a mixed wino-Higgsino 
whereas the second is narrower and corresponds approximately to the pure wino.
The relic density criterion therefore defines one (almost pure wino) 
or two (mixed wino-Higgsino) sides of the two shaded regions, shown in 
Figs.~\ref{fig:MaxRegion_positivemu} and~\ref{fig:MaxRegion_negativemu}, 
corresponding to the pure and mixed wino. 
The dSphs limit defines the other side in the almost pure-wino region, while 
the remaining sides of the mixed wino-Higgsino area are determined by the 
dSphs limit (upper), the condition $|\mu|-M_2=0$, and the antiproton search 
(the arc on the lower side of the mixed region beginning at 
$M_2\simeq 1.9\,$TeV). We recall that  we consider 
the central dSphs limit and those obtained by rescaling up and down by a 
factor of two; the shading in grey within each region 
is used to differentiate between these three choices. 

Next we consider the exclusion limits in the $M_2$ vs.\ $|\mu|-M_2$ plane from 
the 2016 LUX results, which have been obtained as outlined in 
Section~\ref{sec:DD}. As discussed there, the sign of $\mu$ can strongly 
influence the strength of the direct detection limits and consequently the 
allowed parameter space for mixed wino-Higgsino DM. We therefore consider 
the two cases separately.

\subsection{\boldmath $\mu>0$}
\label{sec:positivemu}

%%%%%%%%%%%%%%%%%%%%%%%%%%%%%%%%%%%%%%%%%%%%%%%%%%%%%%%%%%%%%%%%%%%%%%%%%%%%%%
\begin{figure}[!t]
  \centering
  \includegraphics[width=.65\textwidth]{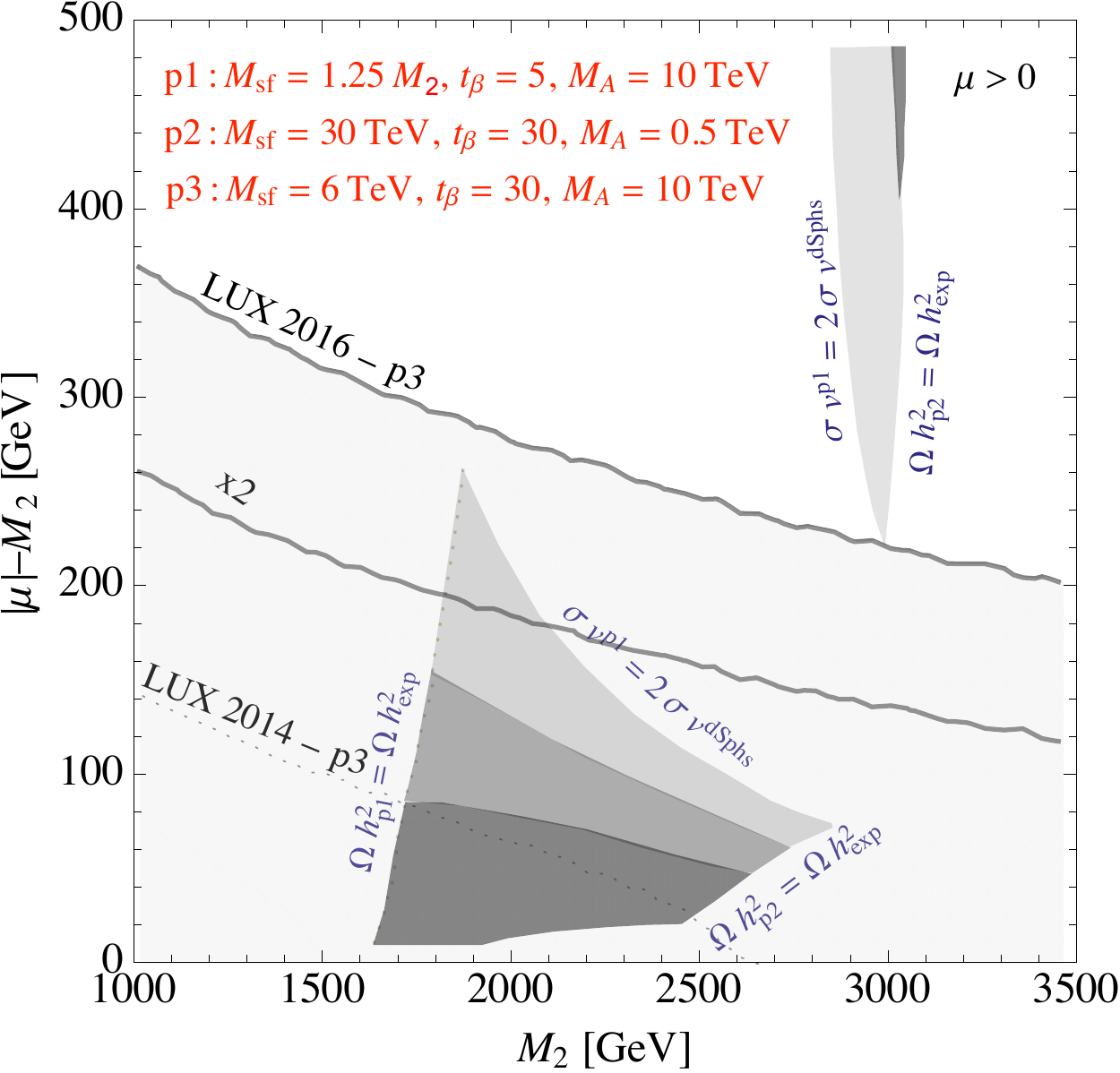}
\caption{Shaded areas denote the maximal region in the $M_2$ vs $|\mu|-M_2$ 
plane for $\mu>0$ where the relic density is as observed and the limit 
from dSphs diffuse gamma searches is respected within parameter 
ranges considered. The darker the grey region, 
the more stringent is the choice of the bound as described in the text. The grey lines mark the \textit{weakest possible} limit
of the region excluded by the 2016 LUX results and the same limit weakened
by a factor of two as indicated. The limit from the previous LUX result is the dotted line. The different bounds are calculated at different parameter 
sets p1, p2 and p3, as indicated.
\label{fig:MaxRegion_positivemu}}
\end{figure}
%%%%%%%%%%%%%%%%%%%%%%%%%%%%%%%%%%%%%%%%%%%%%%%%%%%%%%%%%%%%%%%%%%%%%%%%%%%%%%

Out of the two distinct regions described above, the close-to-pure wino and 
the mixed wino-Higgsino, only the former survives after imposing the direct 
detection constraints, see Fig.~\ref{fig:MaxRegion_positivemu}. 
If conservative assumptions are adopted for direct detection \textit{and} 
dSphs limits a small triangle at the top of the mixed region is still 
allowed. The fact that the direct detection constraints mainly impact the 
mixed rather than the pure wino region was discussed in Section~\ref{sec:DD}, 
and is understood in that the Higgs bosons only couple to mixed 
gaugino-Higgsino neutralinos.

Note that the direct detection limits presented on the plot are for the 
choice of MSSM parameters giving the weakest possible constraints. This is 
possible because the boundaries of the maximal region allowed by indirect 
searches do not depend as strongly on the parameters governing the 
wino-Higgsino mixing as the spin-independent scattering cross section does. 
The only exceptions are the 
boundaries of the mixed-wino region, arising from the relic 
density constraint, which indeed depend strongly on $M_{\rm sf}$.  
However, as varying these boundaries does not significantly change the allowed 
region, since it is mostly in the part excluded by the LUX data, we choose 
to display the LUX bound for a value of $M_{\rm sf}$ different from that 
defining these boundaries. 
Therefore, all in all, the case of the mixed wino-Higgsino with 
$\mu>0$ is verging on being excluded by a combination of direct and indirect 
searches, when imposing that the lightest neutralino accounts for the entire 
thermally produced dark matter density of the Universe. Note, however, 
that the small close-to-pure wino region is not affected by direct detection 
constraints.

\subsection{\boldmath $\mu<0$}
\label{sec:negativemu}

When $\mu<0$ the spin-independent cross section decreases, particularly for 
smaller values of $\tan\beta$. This allows for parameter choices with 
small $|\mu|-M_2$ giving viable neutralino DM, in agreement with the direct 
detection constraint. Indeed, for appropriate parameter choices the direct 
detection limits are too weak to constrain any of the relevant regions 
of the studied parameter space. In particular, the weakest possible limits 
correspond to $M_{\rm sf}=1.25 M_2$, $M_A = 0.5$ TeV and 
$\tan\beta=15$. Note that for $M_A=0.5$ TeV a significantly lower value 
of $\tan\beta$ would be in conflict with constraints from heavy Higgs 
searches at the LHC.

%%%%%%%%%%%%%%%%%%%%%%%%%%%%%%%%%%%%%%%%%%%%%%%%%%%%%%%%%%%%%%%%%%%%%%%%%%%%%%
\begin{figure}[!t]
  \centering
  \includegraphics[width=.65\textwidth]{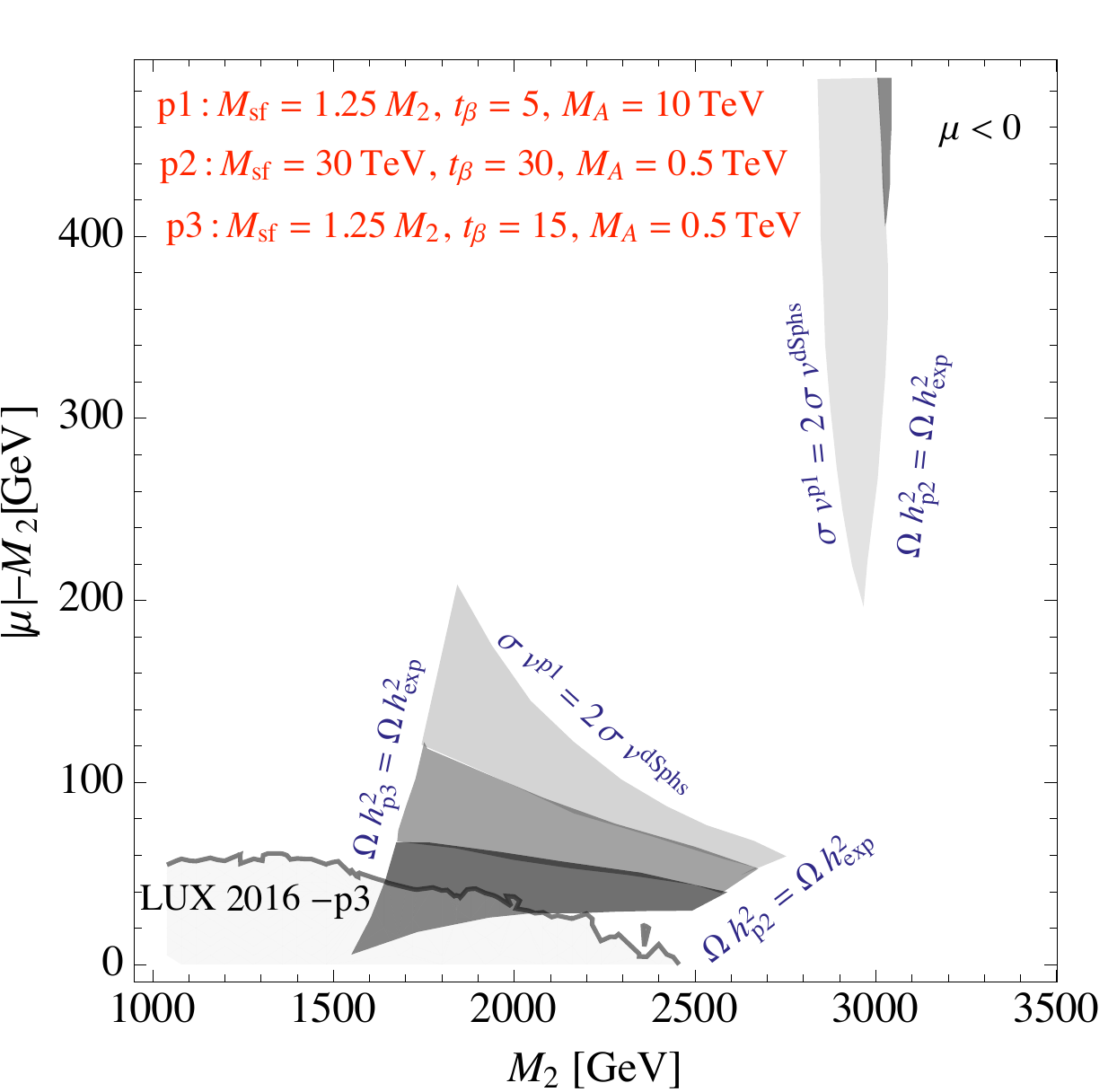}
\caption{Maximal region in the $M_2$ vs $\mu-M_2$ plane for $\mu<0$,
obtained as in Fig.~\ref{fig:MaxRegion_positivemu}. The limit
from the 2016 LUX result weakened by a factor of two 
is not visible within the ranges considered in the
plot. The different bounds are calculated at different parameter 
sets p1, p2 and p3, as indicated. 
\label{fig:MaxRegion_negativemu}}
\end{figure}
%%%%%%%%%%%%%%%%%%%%%%%%%%%%%%%%%%%%%%%%%%%%%%%%%%%%%%%%%%%%%%%%%%%%%%%%%%%%%%

The result of varying $M_A$, $M_{\rm sf}$ and $\tan\beta$ is a sizeable 
mass region for viable mixed-wino dark matter in the MSSM, ranging from 
$M_2=1.6$ to 3 TeV, as shown in Fig.~\ref{fig:MaxRegion_negativemu}. 
The parameter $|\mu|-M_2$ for the 
Higgsino admixture varies from close to 0 GeV to 210 GeV below the 
Sommerfeld resonance, and from 200 GeV upwards above, when the most 
conservative dSphs limit (shown in light grey) is adopted. 

We note that in determining 
the viable mixed-wino parameter region we did not include the 
diffuse gamma-ray and gamma line data from observations of the Galactic 
center, since the more conservative assumption of a cored dark matter 
profile would not provide a further constraint. However, future 
gamma data, in particular CTA observations of the Galactic center, 
are expected to increase the 
sensitivity  to the parameter region in question to the extent 
(\textit{cf.} \cite{Roszkowski:2014iqa}) that either a dominantly-wino 
neutralino dark matter would be seen, or the entire plane 
shown in Fig.~\ref{fig:MaxRegion_negativemu} would be excluded even for 
a cored profile. 

%%%%%%%%%%%%%%%%%%%%%%%%%%%%%%%%%%%%%%%%%%%%%%%%%%%%%%%%%%%%%%%%%%%%%%%%%%%%%%

\section{Conclusions}
\label{sec:conclusions}

This study was motivated by the wish to delineate the allowed parameter (in 
particular mass) range for a wino-like dark matter particle in the MSSM, only 
allowing some mixing with the Higgsino. More generically, this corresponds to 
the case where the  dark matter particle is the lightest state of a heavy 
electroweak triplet with potentially significant doublet admixture and the 
presence of a scalar mediator. The Sommerfeld effect is always important in 
the TeV mass range, where the observed relic density can be attained, and 
has been included in this study extending previous work 
\cite{Beneke:2014gja,Beneke:2014hja,Beneke:2016ync}. Our main results 
are summarized in Figs.~\ref{fig:MaxRegion_positivemu} 
and~\ref{fig:MaxRegion_negativemu}, which show the viable parameter region 
for the dominantly-wino neutralino for the cases $\mu>0$ and $\mu<0$, 
respectively. After imposing the collider and flavour constraints (both 
very weak), we considered the limits from diffuse gamma-rays from the dwarf 
spheroidal galaxies (dSphs), galactic cosmic rays and cosmic microwave 
background an\-iso\-tro\-pies. We also calculated the antiproton flux 
in order to compare with the AMS-02 results. 
The choice of indirect search constraints is 
influenced by the attitude that the fundamental question of the viability 
of wino-like dark matter should be answered by adopting 
conservative assumptions on astrophysical uncertainties. The non-observation 
of an excess of diffuse gamma-rays from dSphs then provides the 
strongest limit. 

It turns out that in addition to these indirect detection bounds, the direct 
detection results have a significant impact on the parameter space, 
particularly for the $\mu>0$ case where the mixed Higgsino-wino region is 
almost ruled out as shown in Fig.~\ref{fig:MaxRegion_positivemu}. 
In the $\mu<0$ case the limits are weaker as seen in 
Fig.~\ref{fig:MaxRegion_negativemu}, and a sizeable viable region remains. 
Note that the region of the $|\mu|-M_2$ vs.~$M_2$ plane 
constrained by direct detection is complementary to that constrained by 
indirect detection. Therefore while for $\mu>0$, (almost) the entire mixed 
region is ruled out, for $\mu<0$ there is a part of parameter space where
$M_2= 1.7-2.7$ TeV which is in complete agreement 
with all current experimental constraints.

Let us conclude by commenting on the limits from line and diffuse 
photon spectra from the Galactic center. If a cusped or 
mildly cored DM profile was assumed, the H.E.S.S. observations of diffuse 
gamma emission \cite{Abdallah::2016jja} would exclude nearly the entire 
parameter 
space considered in this paper, leaving only a very narrow region with close 
to maximal wino-Higgsino mixing. The limits from searches for 
a line-like feature \cite{Abramowski:2013ax} would be even stronger, leaving 
no space for mixed-wino neutralino DM. However, a cored DM profile 
remains a possibility, and hence we did not include the H.E.S.S. results.
In other words, adopting a less conservative approach, one would conclude 
that not only the pure-wino limit of the MSSM, but also the entire 
parameter region of the dominantly-wino neutralino, even with 
very large Higgsino or bino admixture, was in strong tension with the 
indirect searches. Therefore, the forthcoming observations by CTA 
should either discover a signal of or definitively exclude the 
dominantly-wino neutralino.

\subsubsection*{Acknowledgements}

We thank A.~Ibarra for comments on the manuscript, and A.~Goudelis and 
V.~Rentala for helpful discussions.
This work is supported in part by the Gottfried Wilhelm Leibniz programme 
of the Deutsche Forschungsgemeinschaft (DFG) and the Excellence Cluster 
``Origin and Structure of the Universe'' at Technische Universit\"at 
M\"unchen. AH is supported by the University of Oslo through the Strategic 
Dark Matter Initiative (SDI). We further 
gratefully acknowledge that part of this work was performed using the 
facilities of the Computational Center for Particle and Astrophysics (C2PAP) 
of the Excellence Cluster.

%%%%%%%%%%%%%%%%%%%%%%%%%%%%%%%%%%%%%%%%%%%%%%%%%%%%%%%%%%%%%%%%%%%%%%%%
%\bibliography{bib-ID}

\providecommand{\href}[2]{#2}\begingroup\raggedright\endgroup

%%%%%%%%%%%%%%%%%%%%%%%%%%%%%%%%%%%%%%%%%%%%%%%%%%%%%%%%%%%%%%%%%%%%%%%%

\end{document}